\documentclass[prb]{revtex4}
\usepackage{graphicx}
\usepackage{epsfig}
\usepackage{bm}

\begin{document}

\title{Monte Carlo Study of the Nucleation Process during Zeolite
Synthesis}

\author{Minghong G.\ Wu}

\author{and Michael W.\ Deem}

\affiliation{Chemical Engineering Department\\
University of California\\
Los Angeles, CA\ \ 90095-1592}

\newcommand{\rv}[1]{{\mathbf{r}_{#1}}}
\newcommand{\rvs}[1]{{\{\mathbf{r}^{#1}\}}}
\newcommand{\xv}[1]{{\mathbf{x}_{#1}}}
\newcommand{\rvx}[1]{{\mathbf{r}_{\mathrm{#1}}}}
\newcommand{\ns}{{n_{\mathrm{Si}}}}
\newcommand{\no}{{n_{\mathrm{O}}}}
\newcommand{\rvxs}[2]{\{{\mathbf{r}_{\mathrm{Si}}}^{\ns #1}\},\{{\mathbf{r}_{\mathrm{O}}}^{\no #2}\}}
\newcommand{\mus}{{\mu_{\mathrm{Si}}}}
\newcommand{\muo}{{{\mu_{\mathrm{O}}}}}
\newcommand{\Lbs}{{{\Lambda_{\mathrm{Si}}}}}
\newcommand{\Lbo}{{{\Lambda_{\mathrm{O}}}}}
\newcommand{\zs}{{{z_{\mathrm{Si}}}}}
\newcommand{\zo}{{{z_{\mathrm{O}}}}}
\newcommand{\rhos}{{{\rho_{\mathrm{Si}}}}}

\begin{abstract}
An atomic-scale model for silicate solutions is introduced for
investigation of the nucleation process during zeolite
synthesis in the absence of a
structure directing agent.
 Monte Carlo schemes are developed to determine the
equilibrium distribution of silicate cluster sizes within the context
of this model. How the nucleation barrier and critical cluster size
change with Si monomer concentration is discussed. Distance and angle
histograms as well as ring size distributions are calculated and
compared with known zeolite structures. The free energies of critical clusters
are compared with those for small clusters of $\alpha$-quartz.
\end{abstract}

\maketitle

\newpage
\section{Introduction}
\label{sec:introd}
Zeolites are crystalline microporous materials that have found a wide
range of use as catalysts, molecular sieves, and ion
exchangers.\cite{Breck_74,Ruthven_84} A typical example is ZSM-5,
used as a cracking catalyst in the
refinement of crude oil. 
Essentially all refinement of petroleum
occurs through fluid catalytic cracking with a H-ZSM-5 zeolite
catalyst. Essentially all reforming occurs with zeolitic catalysts,
typically H-ZSM-5 or faujasite-type structures. A significant fraction of air
separation into N$_2$ and O$_2$ occurs via a process involving the
zeolite Li-X. Roughly 25\% by weight of the powdered laundry used in
washing machines is the zeolite Na-A, which softens the water for
low-phosphate detergents. Finally, zeolites are widely used in
remediation applications. For example, victims of the
three-mile-island accident were fed ion-exchange zeolites to remove
radioactive traces of $^{127}$Cs$^+$ and $^{90}$Sr$^{2+}$.

Zeolites continue to be synthesized at a furious pace. The major
current limitation of our ability to synthesize zeolites with tailored
properties is an incomplete understanding of the fundamental steps
that occur in zeolite synthesis,\cite{de_Moor_99} which is normally
carried out at hydrothermal conditions.\cite{Helmkamp_95} Important
factors that determine which zeolite will be made include the Si/Al
ratio, the alkalinity, the cation species, and the presence of
templates.\cite{Helmkamp_95,Feijen_94} A significant amount of
experimentation, including NMR spectroscopy, X-ray diffraction, and
neutron scattering, has been done in an effort to characterize the
growth kinetics of
zeolites,\cite{de_Moor_99,Watson_97,III_Burkett,III_Dokter,III_Tang,III_Lechert,III_Chang,III_Hu}
yet the mechanism of nucleation is still not fully understood.  One of
the difficulties arises from the accessible length scale in
experiments. NMR is able to provide information on the 0.5-1.0~nm
scale. Diffraction, on the other hand, is limited to providing
detailed information on the 5~nm and above length scale.  The
nucleation process falls, unfortunately, within this gap and is
difficult to be identified directly. High temperature calorimetry
provides thermodynamic data but no direct structural
information.\cite{Petrovic_93} Recently, it has been possible to
access the nucleation scale with a combination of electron microscopy,
X-ray, and neutron diffraction methods.\cite{de_Moor_99} A working
hypothesis formed is that, at least for template-mediated synthesis of
ZSM-5, the nucleation event involves roughly 8 disordered unit cell
precursors aggregating into a ``primary unit''.
This aggregate forms the nucleation core, to
which additional units are added. It is unclear whether the larger
5-10~nm globule is already crystalline or is amorphous. As the globule
grows, a small crystallite begins to form. This type of nucleation on
the length scale of a few unit cells is consistent with typical
nucleation behavior of other electrolytes in aqueous media. In
particular, the primary unit size for ZSM-5 is roughly 2 unit cells on
a side, consistent with the nucleation scale of 2-3 unit cells for
typical electrolytes condensing in aqueous media. Further support is
provided by the observed primary unit size of 2.6 nm for zeolite beta
(unit cell of 1.3 $\times$ 1.3 $\times$ 2.4 nm)\cite{Tsuji_97} and
1.6 nm for zeolite SOD (unit cell of 0.9 $\times$ 0.9 $\times$ 0.9
nm).\cite{de_Moor_99} These primary units were not detected under
conditions for which no zeolite was formed (i.e., without
template). The concentration of primary units decreased as the initial
stages of nucleation and growth took place. All of these data show the
strong correlation between the presence of the primary units and the
rate of nucleation. This hypothetical mechanism for ZSM-5 synthesis,
however, has not been unambiguously demonstrated to be correct.

The hypothetical zeolite nucleation event involves few enough atoms
that it may be investigated computationally. Molecular dynamics has
been used to examine the early stage of the polymerization process of
silicate ions in aqueous
solution.\cite{Feuston_90,Garofolini_94} This dynamical approach
yields important information of the reaction mechanism, but is limited
to the case where the system is well above the saturation
concentration and the free energy barrier is low. Methods that can
investigate the nucleation event for simple solids have recently been
developed.\cite{Ruiz-Montero_97,ten_Wolde_96} Related methods have
been developed in the context of ion-induced aerosol
nucleation.\cite{KusakaI_98,KusakaII_98} These methods make use of
transition state theory, and so they avoid the time-scale problems
associated with a direct simulation of the nucleation event. 

As a first step towards fundamental understanding of the zeolite nucleation
process, we present here a Monte Carlo study of silicate solutions,
using a model with explicit Si and O atoms.  We aim at constructing an
equilibrium distribution of clusters that provides the free energy
barrier of nucleation.  Novel reactive moves that alter the
connectivities of the silicate clusters, as well as other Monte Carlo
moves, are used to equilibrate the system.  No {\em a priori}
building blocks are assumed in the simulation. Experimentally known
features of aqueous silicate solutions are used to calibrate the
model.  We analyze the structure of the clusters found in the
silicate nucleation process, and we compare the results with known zeolite
structures. 

This article is organized as follows. We describe the model for the
silicate solution in Sec.~\ref{sec:model}.  We describe in
Sec.~\ref{sec:fec} how to compute the equilibrium distribution of
clusters under different experimental conditions. The Monte Carlo
simulation schemes are described in Sec.~\ref{sec:simulation}. Results
are given in Sec.~\ref{sec:res}, and discussions are made in
Sec.~\ref{sec:dis}. Finally, we conclude in Sec.~\ref{sec:conclusion}.

\section{Modeling the Silicate Solution}
\label{sec:model}
A key component of simulation of silicate nucleation is a proper
forcefield to describe the energetics of the silicate species,
hydrogel, cations, and templates. We focus on the pure-silica case for
the nucleation process, and model the silicate clusters by explicit,
charged Si and O atoms. It is advantageous to distinguish the atoms by
their connectivities. Each Si atom is $sp^3$ bonded to four O
atoms. An O atom that is bonded to one O atom is denoted by O$^{(1)}$,
and an O atom that is bonded to two Si atoms is denoted by O$^{(2)}$.
We denote by Si$^{(i)}$ a Si atom that is bonded to
$i$ O$^{(2)}$ atoms and (4-$i$) O$^{(1)}$ atoms and call this an
$i$-connected Si. Neutralizing charges in the solution are treated
implicitly by adding a uniform, neutralizing background charge distribution to
the system. Hydrogen atoms terminating a SiO$^{(1)}$ bond are also
treated in an implicit fashion.

Since the nucleation and crystallization of the silicate species
involves the formation of covalent bonds, we will need to use a
reactive forcefield. Such forcefields have been developed for silicate
species, and they continue to be
developed.\cite{Garofolini_94,WatanabeI_99,Burchart_92,van_Beest_90,Kramer_90,Vashishta_90,Tsuneyuki_88} Here
we use a modified version of the potential proposed by Vashishta et
al.\cite{Vashishta_90} This potential includes pairwise and
three-body interactions between explicit atoms. The two-body part
accounts for the steric repulsion, Coulomb interactions, and
charge-dipole interactions. The explicit form of the two-body
interaction is
\begin{eqnarray}
\label{eqn:2-body}
V_2 & =& \frac{H_{ij}}{r^{\eta_{ij}}}+\frac{q_iq_j}{\epsilon r}
	-\frac{\alpha_i{q_j}^2 +\alpha_j{q_i}^2}{r^4}\exp(-r/r_{4s}).
\end{eqnarray}
Here H$_{ij}$ and $\eta_{ij}$ represent the strengths and exponents of
the steric repulsion. The $q_i$ and $\alpha_i$ are the effective
charge and electronic polarizability of the $i$th ion,
respectively. The decay length $r_{\mathrm{4s}}$ is taken to be 4.43
\AA.\cite{Vashishta_90} The expression for the three-body interaction
is
\begin{equation}
\label{eqn:v3}
V_3 = \left\{\begin{array}{l}
		B_{jik}\exp
		\left[\frac{l}{r_{ij}-r_0}+\frac{l}{r_{ik}-r_0}
		\right]
		(\cos\theta_{jik}-\cos\overline{\theta}_{jik})^2,\; 
		\; r_{ij}<r_0 \mbox{ and } r_{ik}<r_0 \nonumber \\
		0,\; r_{ij}\geq r_0 \mbox{ or } r_{ik}\geq r_0\; ,
	       	\end{array}
	\right.
\end{equation}

where $B_{jik}$ is the strength of the three-body interaction, $l$ is
a constant, $r_0$ is a cutoff distance, $\theta_{jik}$ is the
angle subtended by the vectors $\rv{ij}$ and $\rv{ik}$ with the vertex
at $i$, and $\overline{\theta}_{jik}$ is a constant angle.  The
constants are summarized in Table~\ref{tab:potential}.

One of our modifications of the potential takes into account the
solvent effects, which are modeled by applying a distance-dependent
dielectric constant $\epsilon = 4r$ in
eq.~\ref{eqn:2-body}.\cite{Guenot_92} Our second modification
attempts to incorporate the deprotonation effects.  Since
$q(\mathrm{O}^{(2)})=-0.80$ e, as seen in Table~\ref{tab:potential},
each Si-O bond contributes -0.40 e to the O$^{(2)}$ charge. Silicate
solutions at zeolite synthesis condition have $\mathrm{pH}=11-12$, at which
about half of the Si-O$^{(1)}\cdots$H bonds of monomers are
dissociated.\cite{Iler_79} Assuming an O$\cdots$H pair has a total
charge of -0.40 e, and the de-hydrogenated O$^{(1)}$ has a total
charge of -1.40 e, we assign the average value of -0.90 e to the
O$^{(1)}$ atoms, thus taking into account the terminal hydrogens in
an implicit way.

\section{Determining the Cluster-Size Distribution}
\label{sec:fec}
Direct simulation of the aqueous silicate solution will fail to
observe the nucleation event, because of the extremely low probability
for the relevant fluctuations to occur. Despite this difficulty, it
has been shown that the probability distribution of clusters can be
obtained from simulation of a single cluster in the grand canonical
ensemble.\cite{KusakaI_98,ten_Wolde_99} Such a distribution can be
used to quantify the free energy of clusters and to predict the
nucleation rate from theory.\cite{Ruiz-Montero_97,Frenkel_96}

Consider a system specified by volume $V$, temperature $T$, and the
chemical potentials $\mus$ for Si and $\muo$ for O, where $\mus$ and
$\muo$ are implicitly related to the concentration and pH of the
system. The reference system is taken to be the monomeric
Si(OH)$_2$O$_2^{2-}$ ion system. A uniform, neutralizing background
charge density is added to the reference system so as to make the
system energy finite.\cite{Wu_01}  When the system becomes super-saturated but
confined in the meta-stable region, polymerization of the silicates
leads to larger clusters.  We use the following geometrical criterion
for a cluster:\cite{ten_Wolde_99} a Si-O bond is formed when their
distance is within 2.60 \AA,  each Si atom is bonded to exactly four O
atoms, an O atom can be bonded to either one or two Si atoms, no Si-Si
bonds or O-O bonds can be formed, and a cluster is a set of atoms that are
connected by bonds. We classify a cluster by its numbers of Si and O
atoms, denoted by $\ns$ and $\no$, respectively.  Since nucleation is
a rare event, the population of clusters is very low, and the
interaction between clusters is negligible. In this case, the partition
function of the system can be written as\cite{ten_Wolde_98}
\begin{eqnarray}
\label{eqn:Xi}
\ln\Xi(\mus, \muo, V, T)&= &\sum_{\ns,\no}\exp\left[\beta(\ns\mus+\no\muo)\right]
			\mathrm{Q}_{\ns,\no}\ ,
\end{eqnarray}
where $\beta=1/(k_BT)$ is the reciprocal temperature, and the
partition function for an $\ns,\no$-cluster, $\mathrm{Q}_{\ns,\no}$, is defined as
\begin{eqnarray}
\label{eqn:Zpar}
\mathrm{Q}_{\ns,\no}& =& \frac{V}{\Lbs^{3\ns}\Lbo^{3\no}\ns!\no!}Z_{\ns,\no}\nonumber\\
Z_{\ns,\no}&= &\int d\rvx{Si}^\ns
	     d\rvx{O}^\no
	     \delta(\mathbf{R}_{\mathrm{cm}})
		w(\rvx{Si}^\ns, \rvx{O}^\no)
		\exp[-\beta U(\rvx{Si}^\ns,\rvx{O}^\no)].
\end{eqnarray}
Here $\Lbs$ and $\Lbo$ are the thermal wavelengths for Si and O atoms,
respectively, $\mathbf{R}_{\mathrm{cm}}$ is the center of mass of the
cluster, and the geometrical cutoff function $w(\rvx{Si}^\ns,
\rvx{O}^\no)$ is 1 if $\rvx{Si}^\ns, \rvx{O}^\no$ satisfy the cluster
criterion, and it is 0 otherwise. Note that $w(\rvx{Si}^\ns, \rvx{O}^\no)$
is invariant upon translation of the coordinates. The potential energy
$U(\rvx{Si}^\ns, \rvx{O}^\no)$ accounts for the intra-cluster
interaction. The translational degrees of freedom of the cluster are
integrated out to get $V$. Note that the configurational integral in
eq.~\ref{eqn:Zpar} is independent of $V$ as long as $V$ is larger than
the cluster size.  

The average number of clusters of size $\ns,\ \no$ is
given by\cite{ten_Wolde_98}
\begin{eqnarray}
\label{eqn:N_avg}
\langle N(\ns, \no)\rangle &= &
		\exp\left[\beta(\mus\ns+\muo\no)\right]
		\mathrm{Q}_{\ns,\no}.
\end{eqnarray}
The partition function for a single cluster is given by
\begin{eqnarray}
\Xi_{\mathrm{cluster}} &= & \sum_{\ns,\no}\exp\left[\beta(\ns\mus+\no\muo)\right]
			 \mathrm{Q}_{\ns,\no}.
\end{eqnarray}
That is, the partition function of the system of non-interacting
clusters is the exponential of the partition function
for a single cluster.

We define $\Delta\Omega$ as the excess free energy of
a cluster over an isolated monomer. It is given by
\begin{eqnarray}
\beta \Delta\Omega(\ns, \no, \mus, \muo, T) &= & 
	-\ln\frac{\langle N(\ns, \no)\rangle}{\langle N(1, 4)\rangle}.
\end{eqnarray}
A natural choice of the nucleation coordinate is
$\ns$. Therefore, we are mainly concerned with the dependence of
the free energy on $\ns$. The free energy of the $\ns$-cluster, which is
the union set of $\ns, \no$-clusters with all possible $\no$, is defined
as
\begin{eqnarray}
\label{eqn:ns_cluster}
\beta \Delta\Omega(\ns, \mus, \muo, T) &= & 
	-\ln
	 \left\{\sum_\no
	       \exp\left[-\beta\Delta\Omega(\ns, \no, \mus, \muo, T)\right]
	 \right\}.
\end{eqnarray}
In the simulation only one cluster is present. The variables that
specify the system are $\mus$, $\muo$, and $T$. As shown in
Sec.~\ref{sec:simulation}, an arbitrary system volume $V$ can be used
in simulation, because the particle insertion/removal moves are
proposed according to a geometrical criterion instead of the system
volume.\cite{ten_Wolde_99} To simplify the notation in the following
sections, we define the fugacities $\zs=\exp(\beta\mus)/\Lbs^3$ and
$\zo=\exp(\beta\muo)/\Lbo^3$. Equation~\ref{eqn:N_avg}
can be rewritten as
\begin{eqnarray}
\label{eqn:conc_part}
\frac{\langle N(\ns, \no)\rangle}{V} &= & \frac{\zs^\ns
\zo^\no}{\ns!\no!}Z_{\ns,\no}
\end{eqnarray}

\section{Monte Carlo Schemes}
\label{sec:simulation}
We use several different Monte Carlo moves to sample different degrees
 of freedom in our system. At all times, only a single cluster is
 present in the simulation box. Both the configurational degrees of
 freedoms of the cluster and the numbers of Si and O atoms in the
 cluster are sampled by the Monte Carlo moves. These moves are chosen
 randomly from a set with predetermined probabilities. The move
 frequencies are different for small and large clusters so as to achieve
 effective sampling.  The cluster criterion is enforced at all times
 during the simulation to avoid dissociation of the cluster.

\subsection{Reactive Moves}
The covalent bond formed between Si and O atoms has a magnitude that
is much larger than thermal fluctuations at the synthesis
temperature. Therefore, direct insertion or removal of particles will
fail, due to the rough energy landscape.  Thus, we developed smart
Monte Carlo reactive moves to overcome this problem. The Monte Carlo
reactive moves are divided into three types, the SiO$_n$
insertion/removal move, where $n=2$ or 3, the hydrolysis/condensation
move, and the prune-and-graft move. We demonstrate in
Fig.~\ref{fig:lattice} how the SiO$_n$ insertion/removal move and
hydrolysis/condensation move help the system to equilibrate the
chemical composition $(\ns, \no)$ of the cluster. 
For example, a
SiO$_3$ insertion move brings the cluster from $(\ns=2,\ \no=7)$ to
$(\ns=3,\ \no=10)$,
as shown by the arrow in Fig.~\ref{fig:lattice}.  The prune-and-graft
move does not change $\ns$ and $\no$ and thus is not shown on the
lattice. We here describe the SiO$_2$ insertion/removal moves. Other
moves are described in Appendix~\ref{app:reactive}.

A SiO$_n$ insertion move attempts to graft a SiO$_n$ group onto the
surface of the cluster. A SiO$_n$ removal move attempts to remove a
SiO$_n$ group from the surface of the 
cluster. Figure~\ref{fig:rxnSiO2} illustrates the SiO$_2$
insertion/removal moves.  
A SiO$_2$ insertion move proceeds as
follows:
\begin{enumerate}
\item
Pick randomly a pair of O$^{(1)}$ atoms from a SiO$_2$ insertion list
that is constantly updated during the simulation. The two O atoms must
satisfy certain geometrical criteria, which eliminate geometrically
impossible pairs so as to improve efficiency of the moves.
Only distinct pairs of atoms are included in this list; that is,
only the identity of the pair of atoms, and not the order, is
included in this list.  Consider
the two Si atoms to which this pair of O$^{(1)}$ atoms are
bonded. Mark as group 1 the first Si atom and the O$^{(1)}$ atoms
bonded to it. Mark as group 2 the second Si atom and the O$^{(1)}$
atoms bonded to it.
\item
\label{stp:SiO2_ins}
Generate trial local moves for group 1 and group 2, according to the
O$^{(2)}$ connectivity $i$ of each Si atom. When $i=1$, generate a
random orientation for the SiO$_3$ group. When $i=2$, rotate the
SiO$_2$ around the axis passing through the two O$^{(2)}$ atoms bonded
to the Si atom. When $i=3$, no move is made. In all cases, multiple
trial moves are generated, each move consisting of a modification of
both groups.
One of the trial moves is chosen with a
probability proportional to 
$p(r_{\mathrm{OO}})\exp(-\beta U)/(4 \pi {r_{\mathrm{OO}}}^2)$. Here
$r_{\mathrm{OO}}$ is the distance between the pair of O$^{(1)}$ atoms,
and $p(r_{\mathrm{OO}})\, dr_{\mathrm{OO}} $ is the probability that
two chosen O atoms in a monomer have a distance between
$r_{\mathrm{OO}}$ and $r_{\mathrm{OO}}+ dr_{\mathrm{OO}}$. This
probability is predetermined by a short hybrid Monte Carlo run for a
monomer. The hybrid Monte Carlo will be described in
next section. Here $U$ is the energy of the cluster. The
Rosenbluth factor\cite{Frenkel_96} (the normalization factor) for
this step is calculated and denoted as $w_1$.
\item
Generate a monomeric SiO$_4$ configuration from the Boltzmann
distribution, under the constraint that two of the O atoms are
fixed at the positions of the pair of O$^{(1)}$ atoms. The Fixman
potential associated with this constraint is
zero.\cite{Frenkel_96} Generation of the configuration is
done by a short hybrid Monte Carlo run. We show in
Appendix~\ref{app:p_prob} that the probability for
generating this configuration is
\begin{eqnarray}
\label{eqn:prob_part_c}
p(\mbox{SiO}_4, \mathbf{r}_{\mathrm{OO}}) &= &
	\frac{4\pi{r_{\mathrm{OO}}}^2 w(\rvx{Si}, \rvx{O}^4)\exp(-\beta U_{\mathrm{m}})}
		{p(r_{\mathrm{OO}})Z_{1,4}},
\end{eqnarray}
where $U_{\mathrm{m}}$ represents the energy of the SiO$_4$
monomer. 
\item
Take the SiO$_2$ configuration from the monomer by dropping the two
fixed O atoms, and graft this group onto the O$^{(1)}$ atoms in step~2
in a way so that the O atoms left in the monomer would have fallen on top
of the O$^{(1)}$ atoms. Multiple orientations of the fragment are
generated by rotation of the SiO$_2$ group around the axis passing
through the two O$^{(2)}$ atoms bonded to the Si atom. One of the
orientations is
chosen from a biased probability proportional to
$\exp(-\beta U)$, $U$ being the incremental energy associated with the
SiO$_2$ group. The Rosenbluth factor for this step is calculated and
denoted as $w_2$.
\end{enumerate}
A SiO$_2$ removal move proceeds as follows:
\begin{enumerate}
\item
Pick a Si$^{(2)}$ atom from the cluster at random.  Remove this Si atom
and the two bonded O$^{(1)}$ atoms from the cluster.
\item
Make local moves of the two remaining groups of atoms in the same way
described in step~\ref{stp:SiO2_ins} for the SiO$_2$
insertion, except that $p(r_{\mathrm{OO}})$ is not included in the
biased probability. Calculate the Rosenbluth factor $w_1$.
\end{enumerate}

For both the SiO$_2$ insertion and removal move, a reverse move must
be carried out so as to satisfy detailed
balance.\cite{Frenkel_96} The acceptance probability ratio for these
moves can be derived from the partition function and the
probabilities of proposing the moves.  As shown in Appendix~\ref{app:gcmc},
the acceptance probability is given by
\begin{eqnarray}
\label{eqn:SiO2_acc}
\frac{\mathrm{acc}[\mathrm{o}_{\ns,\no}\rightarrow\mathrm{n}_{\ns+1,\no+2}]}
     {\mathrm{acc}[\mathrm{n}_{\ns+1,\no+2}\rightarrow\mathrm{o}_{\ns,\no}]}
     & = & \zs\zo^2\frac{n(\mathrm{OO})^{\mathrm{(o)}}}
		{n\left(\mathrm{Si}^{(2)}\right)^{\mathrm{(n)}}}
	   \frac{Z_{1,4}w_2^{\mathrm{(n)}}/k_2}{
            2! \exp(-\beta U_{\mathrm{m}})}
		\frac{w_1^{\mathrm{(n)}}}{w_1^{\mathrm{(o)}}}.
\end{eqnarray}
Here $n(\mathrm{OO})$ is the number of qualified O$^{(1)}$ pairs for
the insertion move, $n\left(\mathrm{Si}^{(2)}\right)$ is the
number of qualified Si$^{(2)}$ atoms for the removal move, and $k_2$
is the number of trial moves for the SiO$_2$ insertion.

\subsection{Hybrid Monte Carlo}
\label{sec:hbmc}
The hybrid Monte Carlo scheme\cite{Duane_87} has the basic idea that
one can use molecular dynamics to generate global trial Monte Carlo
moves. It has the advantage that a time step too big for
molecular dynamics can be used without causing numerical
instability. The molecular dynamics algorithm used to generate trial
moves must be area preserving and time reversible to guarantee
detailed balance.\cite{Mehlig_92} Our hybrid Monte Carlo move
equilibrates mainly the strong bond vibration and bending motions with short
time scales.  The time step for the molecular dynamics in hybrid Monte Carlo is
chosen to be around 5 fs. To initialize each move, a Gaussian velocity
distribution for the atoms is generated. The molecular dynamics then
proceeds for a fixed number of time steps, and a Metropolis
criterion is used to accept or reject such a move.

\subsection{CBMC Moves} 
\label{sec:cbmc}
The configurational bias Monte Carlo method (CBMC), first developed by
Frenkel, Smit, and de Pablo,\cite{Frenkel_96,dePablo92} successfully
samples complex energy landscapes by using local information when
proposing moves.  Our CBMC scheme proceeds by selecting randomly a
Si-O$^{(2)}$ bond. If the cluster does not break into two independent
clusters when this bond is removed, then we reject this move. This test
is performed for successive bonds to be regrown. The two clusters so
generated are reconnected by regrowth of the deleted fragments along the O$^{(2)}\rightarrow$ Si direction unit by unit,
beginning from the SiO$_3$ group bonded to the O$^{(2)}$. Each SiO$_3$
group is treated as a rigid unit in the regrowth. The reverse move is
performed so as to satisfy detailed balance. This type of move
equilibrates the Si-O-Si bending angles and the torsional angles
around the Si-O bonds. All the bond distances are preserved.

\subsection{Window Sampling}
Although the cluster distribution can be obtained in principle from a
direct simulation in the grand canonical ensemble, the large free
energy barrier to nucleation makes sampling of large clusters infrequent and
statistically poor. The cluster distribution is best calculated by
patching together the curves from several simulations sampling
different ranges of 
$\ns$. This scheme is called window
sampling.\cite{Frenkel_96}  A biasing potential of this form is added
to the system so as to sample different clusters with roughly equal
probability. We chose a linear function of $\ns$ as a biasing potential,
with the slope determined from a short trial Monte Carlo run. This
bias was corrected for when computing the averages.

\section{Results}
\label{sec:res}
\subsection{Modeling the Synthesis Condition}
We performed the cluster simulations in the grand canonical ensemble
specified by $\zs$, $\zo$, and $T=430$~K, a typical temperature for
zeolite synthesis. To estimate the fugacities of the Si and O atoms, we
make use of the experimental data by the requirement that the
equilibrium constant between monomers and dimers in simulation matches
that of the synthesis condition. We assume that at $\mathrm{pH}=12$
the dominant monomeric and dimeric ions are Si(OH)$_2$O$_2^{2-}$ and
Si$_2$(OH)$_4$O$_3^{2-}$, respectively, and look for the equilibrium
constant of this reaction
\begin{equation}
\label{eqn:dimerize}
\mathrm {2\, Si(OH)_2 O_2^{2-} + H_2 O \rightleftharpoons
Si_2(OH)_4O_3^{2-}
	+ 2\, OH^-}.
\end{equation}
This equilibrium constant is estimated from several reactions,
including the dimerization of neutral silicate
species,\cite{Sefcik_97,Cary_82} the dissociation of
silicates,\cite{Iler_79,Sefcik_97} and the dissociation of
water.\cite{Tanger_89} The estimated value of the dimerization
constant is $10^{-3.2}$ M (mol/l). Since the dissociation constant of
water at 430~K is about $10^{-11.7}$ M$^2$, and the pH is 12, the
$\mathrm{OH^-}$ concentration is estimated to be
$10^{-11.7+12}=2.0$~M. Therefore,
\begin{eqnarray}
\label{eqn:dimer_c_exp}
\frac{c_{\mathrm{d}}}{{c_{\mathrm{m}}}^2}&= &
		\frac{10^{-3.2}\ \mathrm{M}}{{c_{\mathrm{OH^-}}}^2}
		=  10^{-3.8}\ \mathrm{M}^{-1},
\end{eqnarray}
where $c_{\mathrm{d}}$ and $c_{\mathrm{m}}$ are the concentrations of
the dimer and the monomer in units of M, respectively.  On the other
hand, from eq.~\ref{eqn:conc_part} we can deduce
\begin{eqnarray}
\label{eqn:dimer_c_sim}
\frac{c_{\mathrm{d}}}{{c_{\mathrm{m}}}^2}&= & 
\frac{1!^2\, 4!^2}{2!\, 7!}\frac{Z_{2,7}}{\zo
{Z_{1,4}}^2}\ (\mathrm{\AA}^3/\mathrm{no.}),
\end{eqnarray}
where $c_{\mathrm{d}}$ and $c_{\mathrm{m}}$ are the
concentrations of the dimer and the monomer in units of
$(\mathrm{no.}/\mathrm{\AA}^3)$, respectively.  Using
$Z_{1,4}=\exp(311.12)\ \mathrm{\AA}^{12}$
together with a short Monte Carlo run to
determine the left hand side of eq.~\ref{eqn:dimer_c_sim},
we find $Z_{2,7}=\exp(623.00)\ \mathrm{\AA}^{24}$
(see Appendix~\ref{app:monomer_p}).
 Equating
eq.~\ref{eqn:dimer_c_exp} and eq.~\ref{eqn:dimer_c_sim}, we obtain
$\zo=0.465\ \mathrm{\AA}^{-3}$, which we use in all our
simulations.
Actually, simulations directly at this value of $\zo$ are rather difficult
to equilibrate;  we perform simulations at a somewhat higher
$\zo$ and reweight the data to this value.
 The fugacity of Si is related to the reference system Si
concentration by $\rhos=\zs\zo^4 Z_{1,4}/(1!4!)\ (\mathrm{
no.}/\mathrm{\AA}^3)$. To estimate the Si concentration, we consider
a typical silicate solution including 8\% of SiO$_2$ and 92\% of
water. Assuming that about 30\% of the silica dissolve in water and
that the density of the solution is 1 (g/cm$^3$), we find an estimated
Si concentration of $0.4$ M. The existence of a gel phase may cause a
range of non-uniform local concentrations, and so we actually explore
a range of Si concentrations.

Window sampling of different cluster sizes, $\ns$, is performed from
$\ns=1$ to $\ns=200$. Each window covers 4 or 5 values of $\ns$. Each
run starts with a configuration taken from the preceding run. Five
thousand Monte Carlo cycles were performed for $1\leq\ns<50$, and at
least 10$^4$ Monte Carlo cycles were performed for $50\leq\ns\leq
200$. Each cycle consists of 50 Monte Carlo moves. The probability
distributions obtained from these windows were patched together to get
the free energies of cluster formation. Simulations with different initial
configurations were used as a check to ensure the system is well
equilibrated.  No hysteresis was observed within the range of this
study.

\subsection{Structure Analysis}


 Common experimentally identified silicate species such as 4-ring
($\ns=4$) and double 4-ring (cubic octamer,
$\ns=8$)\cite{Kinrade_98,Hendricks_91} were found in our
simulation at small cluster sizes. Figure~\ref{fig:connect} shows the
average number of silicon atoms as a function of connectivities for
different values of $\ns$.
 The 1-connected Si atoms are hardly
observed for $\ns>25$, indicating that the cluster is relatively
compact. Figure~\ref{fig:ns_77} shows a snapshot of a $\ns=77$
cluster, which is nearly spherical.

The observed bond length of Si-O in our simulation at 430~K is
$1.67\pm 0.04$ \AA.  Figure~\ref{fig:T-T} shows the probability of the
nearest neighbor Si-Si distance for a large cluster with
$\ns=194$. 
Only the distances between 4-connected Si atoms were
calculated. As a comparison, we computed the probabilities for the T-T
distance, T being Si or Al, for both the 28 all-Si zeolites and the 68
aluminosilicate zeolites in the International Zeolite Association
database.\cite{IZA_web} The simulation distribution is peaked around
$3.05$~\AA, while the distributions from known zeolite structures are peaked
around $3.1$~\AA. Interestingly, the distribution produced from our
all-Si atomic model is closer to the distribution from all-Si zeolite
structures. We also calculated the probability of the Si-Si-Si angle,
shown in Fig.~\ref{fig:T-T-T}. 
Also shown in the same figure are the
distributions for the 24 all-Si zeolite structures and the 68 aluminosilicate
zeolite structures. The all-Si distribution displays a sharp peak
around $110^\circ$, which is close to the value for a perfect
tetrahedron.  The distribution has a shoulder around $90^\circ$, which
implies an abundance of 4-rings. The small peak near $60^\circ$ in the
simulation data comes from 3-rings, which are occasionally found in
zeolite structures. The asymmetry is these distributions in due to
steric repulsion for small ring sizes.

We calculated the ring size distribution for $\ns=77$ and $\ns=200$ by
counting in the cluster the fundamental rings, i.e., those rings that cannot
be divided into two smaller ones.\cite{Stixrude_90} As shown in
Fig.~\ref{fig:ring_size}, the ring size is mostly between 4 and 8,
consistent with the structures of SiO$_2$ compounds.  Note that in
crystalline zeolite structures there are several non-planar,
fundamental rings
that are not normally considered to be rings, \emph{e.g.}\ there
is a fundamental ring with 12 silicons that encircles the
sodalite cage in the zeolite SOD.  
In the histogram of crystalline zeolite rings, therefore,
we have eliminated all fundamental rings that are not sufficiently planar

 Larger rings
appear when the cluster grows, and for $\ns=200$ we observed up to
11-rings. The smooth distribution of ring sizes suggests the cluster is
amorphous.  Interestingly, 3-rings appear only for the 
larger clusters.  As shown in Fig.~\ref{fig:3ring},
the 3-rings are mainly distributed near the surface of the
cluster.  This result is in accord with experiments on
amorphous silica, where 3-rings are observed only on the relatively
flat surfaces of the silica.\cite{Brinker}

\subsection{Free Energy of Clusters}
Figure~\ref{fig:cl_g} shows the cluster free energy at a set of
different Si monomer concentrations. 
Table~\ref{tab:barrier} lists the
estimated nucleation barrier and the corresponding critical cluster
size.
Clearly, the nucleation barrier and the critical cluster size
increase as the Si monomer concentration decreases.  For $\rhos=0.017$ M the
cluster
 free energy of cluster does not pass through a maximum within the range of
this study. The positive slope suggests that the monomer phase is 
stable at this concentration. The cluster free energy becomes almost linear
for $\ns>100$ at this density, suggesting that the bulk energy term
starts to dominate the cluster energy.

To investigate the relative stability of the clusters predicted by our
forcefield, we calculate the Helmholtz free energy at 430~K for both
the $\alpha$-quartz and the MFI framework structure using thermodynamic
integration.\cite{Frenkel_96} The reference state is taken to be the
Einstein crystal, whose free energy can be calculated
analytically.\cite{Polson_00} Our calculation shows crystalline quartz is
more stable in solution than the MFI structure by 14~kcal/mol, while solution
calorimetry results give an enthalpy difference of 1.5~kcal/mol at 298
$^\circ$K.\cite{Petrovic_93} This seems to indicate that our
forcefield over-stabilizes quartz. Such a tendency has been observed
in other forcefields for Si, O systems.\cite{Piccione_00}  Although
quartz is the thermodynamically most stable structure in air at this
temperature, this may not be true in the solution or for small
crystallite sizes. Cations and structure-directing molecules affect
the surface energy and pore size distribution, and thus change the
equilibrium distribution of
clusters.\cite{III_Burkett,Kinrade_98} Therefore, the composite
formed by the structure-directing molecules and ZSM-5 may be more
stable than quartz under aqueous synthesis conditions. From our simulation we
did not observe quartz-like clusters for $\ns<200$. To see if quartz
clusters are less stable than the clusters observed from simulation,
we generated a few small clusters from the crystal structure of
quartz and calculated the free energy in solution by thermodynamic
integration at $\rhos=0.33$ M. There is no unique way of choosing the
atoms that constitute the cluster, and we choose the atoms such that
the cluster is nearly spherical and has no 1-connected Si atoms. As
shown in Fig.~\ref{fig:cl_g}, the calculated free energy of quartz
clusters are all above the curve corresponding to $\rhos=0.33$ M. 
This
suggests the formation of a quartz nuclei may require a higher
activation energy at this condition.  Scattering in the data comes
mainly from our arbitrary choice of clusters. When the cluster is
large enough so that the bulk term becomes dominant, we calculate that
the free energy
of quartz decreases with a slope of -12.8/per Si atom,
which is calculated from the free energy of a quartz crystal and the
fugacities of Si and O atoms. The estimated asymptotic slope of the
amorphous clusters identified in the simulation at $\rhos=0.33$ M is
-2.8/per Si atom. Thus, we expect quartz to become relatively more
stable for large clusters.


\section{Discussion}
\label{sec:dis}
Our analysis of the clusters suggests that more open, amorphous
structures are favored at low values of $\ns$. The Ostwald ripening
rule says that the condensed phase that forms during nucleation is not
necessarily the thermodynamically most stable phase, but rather the
phase that has the lowest nucleation
barrier.\cite{ten_Wolde_96,Zones_01} It is possible that the quartz
phase has a much higher nucleation barrier and is therefore favored
only at large values of $\ns$.

It is expected that both the concentration and pH value have an
influence on the critical cluster size and the nucleation barrier. As
shown in Fig.~\ref{fig:cl_g}, both the critical cluster size and the
nucleation barrier decrease as the Si monomer concentration
increases. The pH value affects the critical cluster size and the
nucleation barrier through the oxygen chemical potential, $\muo$. This
effect can be investigated by reweighting the cluster distribution at
different values of $\muo$, as shown in Fig.~\ref{fig:cl_g_pH}. Here
the Si monomer concentration is chosen to be 0.33~M in all
distributions.  The sensitivity of the nucleation barrier to the 
pH value is characteristic of the sensitivity of zeolite synthesis
to solution conditions.
 A decrease of pH would shift the equilibrium of
eq.~\ref{eqn:dimerize} to the right and lead to lower nucleation
barriers and smaller critical clusters. This does not mean that
lowering the pH value always leads to faster crystallization, since
other factors such as insufficient supply of monomers may prevent the
formation of the nucleus of the crystal. Indeed, the mechanism of
nucleation for zeolites seems to operate only at high pH, where the
critical cluster size is on the order of 50 silicons.

We make a rough estimate of the nucleation rate in this
system. The steady state nucleation rate, $I^{\mathrm{s}}$, is given
from classical nucleation theory as\cite{Kelton_91}
\begin{eqnarray}
\label{eqn:nu_rate}
I^{\mathrm{s}} &= & \frac{24 D {n^*}^{2/3} \rhos N_{\mathrm{A}}}{\lambda^2}
		 \left(\frac{\beta \vert \Delta\mu \vert}{6\pi n^*}\right)^{1/2}
		\exp(-\beta\Delta\Omega^*) \ .
\end{eqnarray}
Here $D$ denotes the diffusion coefficient of the monomers in the
solution, $\lambda$ stands for the atomic jump distance in the
solution, $n^*$ is the critical cluster size, $\Delta\mu$ is the free
energy difference between the monomer phase and the condensed phase,
$\Delta\Omega^*$ is the nucleation barrier, and $N_\mathrm{A}$ is
Avagadro's number. We pick $\rhos=0.33$~M and use the simulation
results $n^*=32$, $\beta \vert \Delta\mu \vert =2.8$, and $\beta
\Delta\Omega^*=91.6$. The diffusion coefficient $D$ is estimated to be
$10^{-5}$~cm$^2$/s, and the atomic jump distance $\lambda$ is
estimated to be 1~\AA.  Substituting these values into
eq.~\ref{eqn:nu_rate}, we obtain $I^{\mathrm{s}}=5.5\times
10^{-32}$~(s$^{-1}\cdot$\AA$^{-3}$). We calculate the rate of
nucleation events under typical non-templated synthesis conditions of
a solution volume of about 1000 cm$^3$ and roughly 1 hour. The
probability of nucleation predicted for such an experiment is
$1.9$. This rough estimate is closer than is typical for
comparisons of homogeneous nucleation theory and
experiment.\cite{Kelton_91,Auer_01} While silicate nucleation is
often thought to be a strongly heterogeneous process, the present
results suggest that the degree of heterogeneity may not be as strong
as commonly thought.  Indeed, many zeolite species can be synthesized
without the use of a structure directing agent, and it may be that
local fluctuations in solution conditions can be responsible for
enhancing the rate of nucleation.  Even
under the assumption of a perfectly homogeneous nucleation event,
discrepancies between our estimate and the true rate may result from
use of a simplified forcefield, uncertainties in the nucleation
barrier, and assumptions in classical nucleation theory.

Large rings with more than 10 constituent Si atoms are found in a few
zeolites. These rings form one or two dimensional pores that are
responsible for unique adsorption and catalytic
selectivities. However, synthesis of zeolites with large rings is
difficult and almost always requires the use of structure directing
molecules such as tetrapropyl ammonium ions. The low population of
large rings in the clusters observed in the present simulations,
therefore, can be attributed partly to the absence of structure
directing molecules.

The synthesis condition modeled in this work strongly favors clusters
with higher connectivities. It is observed during the simulation that
Si and O atoms on the surface rearrange themselves to form bonds and
thus minimize the cluster energy. As a result, 1-connected Si
atoms are hardly observed for clusters larger than 25 Si atoms, as
shown in Fig.~\ref{fig:connect}. For very large clusters we expect
that almost all Si atoms are 4-connected.

The nearest neighbor Si-Si distance observed in the cluster, when
compared with the distribution calculated from zeolites, tends to be
more condensed. Nevertheless, it displays a broad distribution, and
therefore large distances that can lead to expanded structures are
also found. The probability distribution for the Si-Si-Si angle
observed in the cluster is closer to the distribution of the 68 aluminosilicate
zeolites.  Aluminum atoms are known to increase disorder in the
crystal structures. For example, highly strained 3-rings are found
only in zeolites having aluminum or other non-silicon
 elements. Such disorder is
also found in the finite size clusters studied here,
 as indicated by the presence
of smaller rings such as 3-rings and 4-rings. Note that the
experimental peaks corresponding to 4-rings and perfect tetrahedral
order are reproduced in the simulation.

\section{Conclusion}
\label{sec:conclusion}
We have constructed an atomic-scale model for the silicate solution
during zeolite synthesis in the absence of structure
directing agents. Novel Monte Carlo moves were
developed to sample the equilibrium distribution of
silicate clusters. The nucleation barrier is estimated to be on the
order of $10^2\ k_BT$, and the critical cluster size is estimated to
be between $25$ and $50$ silicons. Interestingly, the nucleation
behavior is rather sensitive to the Si concentration in and the pH of the
solution, in
accord with experimental experience. Structural analysis shows the
results reproduce physical properties of condensed phases of
silica, such as the distance distribution and topological features. The smooth
distribution of ring sizes suggests the cluster is amorphous up to
$\ns=200$.  Future work is planned to take into account the effect of
ions and structure-directing molecules on the energetics of the
clusters.

\section*{Acknowledgment}
We thank Mark E. Davis of Caltech and Stacey I. Zones of Chevron
Research Technology Company for helpful discussion. We also thank
Pieter ten Wolde for discussion about the free energy of cluster
formation. This
research was supported by Donors of the Petroleum Research Fund, the
UC Energy Institute, Chevron Research and Technology Company, and the
National Science Foundation.

\appendix
\section{Generating a Monomer Configuration
 with a Fixed $\mathbf{r}_{\mathrm{OO}}$}
\label{app:p_prob}
We consider the configurational integral of a monomer with two of its
O atoms fixed. Let $\mathbf{r}_{\mathrm{OO}}$ be the vector that spans
the two fixed O positions. Let $p(r_{\mathrm{OO}})\,dr_{\mathrm{OO}}$
be the probability that two chosen O atoms of a monomer have a
distance between $r_{\mathrm{OO}}$ and $r_{\mathrm{OO}}+
dr_{\mathrm{OO}}$. Then
\begin{eqnarray}
p(r_{\mathrm{OO}})\,dr_{\mathrm{OO}}& = & \frac{4\pi{r_{\mathrm{OO}}}^2\,
dr_{\mathrm{OO}}}{Z_{1,4}}\int d\rvx{Si}
	     d\rvx{O}^4
	     \delta(\mathbf{R}_{\mathrm{cm}})
		w(\rvx{Si}^\ns,	\rvx{O}^\no)
\nonumber\\
& &\times\delta(\rvx{O_2}-\rvx{O_1}-\rvx{\mathrm{OO}})\exp[-\beta U(\rvx{Si},\rvx{O}^4)],
\end{eqnarray}
where the integral at the right hand side is the constrained
configurational integral of our interest. Therefore, the probability
for generating the constrained monomer configuration, as given in
eq.~\ref{eqn:prob_part_c}, is
\begin{eqnarray}
\label{eqn:SiO4_OO}
p(\mbox{SiO}_4, \mathbf{r}_{\mathrm{OO}}) &= &
\frac{w(\rvx{Si},\rvx{O}^4)\exp(-\beta U)}
		{\int d\rvx{Si}d\rvx{O}^4
	     \delta(\mathbf{R}_{\mathrm{cm}})
		\delta(\rvx{O_2}-\rvx{O_1}-\rvx{\mathrm{OO}})
		w(\rvx{Si}, \rvx{O}^4)
\exp[-\beta U(\rvx{Si},\rvx{O}^4)]}\nonumber\\
&= & \frac{4\pi{r_{\mathrm{OO}}}^2 w(\rvx{Si},\rvx{O}^4)\exp(-\beta U)}
		{p(r_{\mathrm{OO}})Z_{1,4}}.
\end{eqnarray}
The probability
$p(r_{\mathrm{OO}})$ is predetermined with a short hybrid Monte Carlo
run.

\section{The Acceptance Ratio in Grand 
Canonical Monte Carlo}
\label{app:gcmc}
Consider a simple, monatomic gas in a grand canonical ensemble. 
The grand canonical partition function of the system is
\begin{eqnarray}
\label{eqn:gcmc_p}
\Xi &= &\sum_N \frac{z^N}{N!}\int d\rv{}^N \exp\left[-\beta U(\rv{}^N)\right],
\end{eqnarray}
where $z=\exp(\beta\mu)/\Lambda^3$ is the fugacity, $N$ is the number
of particles, $\beta$ is the reciprocal temperature, and $U$ is the
energy of the system. The $N!$ comes from the
indistinguishability of the particles, and it is used here to remove
overcounting of the states. A more convenient way to think about the
density of the states is to rewrite eq.~\ref{eqn:gcmc_p} as
\begin{eqnarray}
\label{eqn:gcmc_pn}
\Xi &= &\sum_N z^N\int d\rvs{N} \exp\left[-\beta
				U\left(\rvs{N}\right)\right],
\end{eqnarray}
where $\rvs{N}$ is the set of coordinates without individual
labeling of atoms. Thus, the probability of a state is given by
\begin{eqnarray}
\rho\left(\rvs{N}\right) &= & \frac{z^N}{\Xi}\exp\left[-\beta
U\left(\rvs{N}\right)\right].
\end{eqnarray}

The detailed balance condition says we must accept the particle insertion and
particle removal moves by the following criterion:
\begin{eqnarray}
\label{eqn:detailed_balance_s}
\frac{\mathrm{acc}(\rvs{N}\rightarrow\rvs{N+1})}
     {\mathrm{acc}(\rvs{N+1}\rightarrow\rvs{N})} &= &
	\frac{\rho(\rvs{N+1})}{\rho(\rvs{N})}
	\frac{\alpha(\rvs{N+1} \rightarrow\rvs{N})}
	     {\alpha(\rvs{N} \rightarrow\rvs{N+1})},
\end{eqnarray}
where $\alpha$ denotes probability of proposing a move, and $\mathrm{acc}$
denotes probability of accepting the proposed move. Let $p$ be the
probability of attempting a particle insertion or a particle
removal move. We choose to perform an insertion or a removal move with equal probability.
For the insertion move we insert a particle in the system volume $V$ at
random. Thus,
\begin{eqnarray}
\label{eqn:alpha_ins}
\alpha\left(\rvs{N}\rightarrow\rvs{N+1}\right) &= & p\frac{1}{2} \frac{1}{V},
\end{eqnarray}
For the removal move we choose at random a particle to be removed, and
\begin{eqnarray}
\label{eqn:alpha_rem}
\alpha(\rvs{N+1}\rightarrow\rvs{N}) &= & p\frac{1}{2} \frac{1}{N+1},
\end{eqnarray}
Substituting eqs.~\ref{eqn:alpha_ins} and~\ref{eqn:alpha_rem} into
eq.~\ref{eqn:detailed_balance_s}, we get
\begin{eqnarray}
\frac{\mathrm{acc}(\rvs{N}\rightarrow \rvs{N+1})}
     {\mathrm{acc}(\rvs{N+1}\rightarrow \rvs{N})} &= &
	\frac{z V}{N+1}\exp
\left[-\beta\left[U\left(\rvs{N+1}\right)-U\left(\rvs{N}\right)
			\right] \right] .
\end{eqnarray}
This is the usual acceptance ratio.\cite{Frenkel_96,Allen}

In the present case, we have a two component system, so for the fixed
center of mass single cluster
\begin{eqnarray}
\label{eqn:p_states}
\rho(\rvxs{}{}) &= & \frac{\zs^\ns\zo^\no}{\Xi_{\mathrm{cluster}}}
	\exp\left[-\beta U\left(\rvxs{}{}\right)\right].
\end{eqnarray}
For insertion or removal of a SiO$_2$ fragment, detailed balance reads
\begin{eqnarray}
\label{eqn:detailed_balance}
\frac{\mathrm{acc}[\mathrm{o}_{\ns,\no}\rightarrow\mathrm{n}_{\ns+1,\no+2}]}
     {\mathrm{acc}[\mathrm{n}_{\ns+1,\no+2}\rightarrow\mathrm{o}_{\ns,\no}]}
&= &
	\frac{\rho\left(\rvxs{+1}{+2}\right)}{\rho\left(\rvxs{}{}\right)}\nonumber\\
& &\times
\frac{\alpha\left[\mathrm{n}_{\ns+1,\no+2}\rightarrow\mathrm{o}_{\ns,\no}\right]}
     {\alpha\left[\mathrm{o}_{\ns,\no}\rightarrow\mathrm{n}_{\ns+1,\no+2}\right]}.
\end{eqnarray}
Here $\mathrm{o}_{\ns,\no}$ represents the old state with particle
positions $(\rvxs{}{})$, and $\mathrm{n}_{\ns+1,\no+2}$ represents
the new state with particle positions $(\rvxs{+1}{+2})$.  Let $k_1$ be
the number of trial moves generated for the two Si groups, and let
$k_2$ be the number of trial moves generated for the SiO$_2$
fragment. We choose $k_1=50$ and $k_2=50$. The probability of
generating a SiO$_2$ insertion move, $\alpha[{\mathrm
o}_{\ns,\no}\rightarrow\mathrm{n}_{\ns+1,\no+2}]$, is the product of
probabilities of each selection process as described in the
text. Trial moves for the SiO$_2$ removal move must be generated so as
to calculate the acceptance probability. Detailed balance in this case
describes the transition probabilities between two states that are not
only characterized by their particle coordinates but also by the trial
moves generated in the intermediate steps.  We impose the
super-detailed balance condition.\cite{Frenkel_96} It follows that
\begin{eqnarray}
\label{eqn:SiO2_ins_gen}
\alpha[\mathrm{o}_{\ns,\no}\rightarrow\mathrm{n}_{\ns+1,\no+2}] 
&= & \frac{1}{n(\mathrm{OO})^{\mathrm{(o)}}}
 \frac{1}{2}
\frac{k_1!}{{C_1}^{k_1}}
\frac{p(r_{\mathrm{OO}})\exp\left(-\beta{U_1}^{\mathrm{(n)}}\right)}
{4 \pi {r_\mathrm{OO}}^2 w_1^{\mathrm{(n)}}}
\nonumber\\
&&
\times 2\pi p(\mbox{SiO}_4, \mathbf{r}_{\mathrm{OO}})
\frac{k_2!}{(2\pi)^{k_2}}\frac{2! \exp\left(-\beta{U_2}^{\mathrm{(n)}}\right)}{
w_2^{\mathrm{(n)}}} \nonumber\\
&&\times\frac{(k_1-1)!}{{C_1}^{k_1-1}}
\nonumber\\
&= &
\frac{1}{n(\mathrm{OO})^{\mathrm{(o)}}}
 \frac{
w(\rvx{Si},\rvx{O}^4)}{Z_{1,4}}
\nonumber\\
& &\times
\frac{\exp\left[-\beta(U_m+{U_1}^{\mathrm{(n)}}+ {U_2}^{\mathrm{(n)}})\right]}
  {w_1^{\mathrm{(n)}}w_2^{\mathrm{(n)}}}\nonumber\\
& &\times \frac{k_1! (k_1-1)!}{{C_1}^{2k_1-1}}  \frac{k_2!}{(2\pi)^{k_2-1}},
\end{eqnarray}
${U_1}^{\mathrm{(n)}}$ and ${U_2}^{\mathrm{(n)}}$ denote the
energy of the two chosen Si groups and the inserted SiO$_2$ group,
respectively, and $C_1$ is a constant that is related to the phase
space volume generating trial moves for the two Si groups. The value
of $C_1$ is irrelevant because it cancels in the acceptance ratio.
The factor of $1/2$ accounts for the two possible choices for labeling
groups 1 and 2.
The factor $2\pi$ before $p(\mbox{SiO}_4, \mathbf{r}_{\mathrm{OO}})$
accounts for the fact the rotational angle of the SiO$_2$ fragment has
not been chosen yet; equivalently, this factor accounts for the
values of the rotational angle that contribute to the flux.
The factor of $2!$ accounts for the number of indistinguishable 
SiO$_2$ configurations
contributing to the flux.
The Rosenbluth factors are defined by $w_1^{\mathrm{(n)}}=
\sum_{k=0}^{k_1} p({r_{\mathrm{OO}}}^{(k)})\exp\left(-\beta
U_1^{(k)}\right)/ (4 \pi {r_\mathrm{OO}}^2)$ 
and $w_2 = \sum_{k=0}^{k_2} \exp(-\beta
U_2^{(k)})$.  Note that a biased insertion volume instead of the system
volume is used. The $k_1!$ and $k_2!$ account for the number of
indistinguishable 
trial moves that contribute to the flux.
 The term $(k_1-1)!/{C_1}^{k_1-1}$ at the right end of
the first equality of eq.~\ref{eqn:SiO2_ins_gen} accounts for the
probability of generating $k_1-1$ trial moves for the corresponding
SiO$_2$ removal move. These $k_1-1$ trial moves combined with the
configuration chosen in the insertion move are used to calculate the
Rosenbluth factor $w_1^{(\mathrm{o})}$, which is described
below. Inserting eq.~\ref{eqn:SiO4_OO} into the right hand yields the
second equality. Since we are considering transitions between accessible
configurations, the geometrical cutoff function
$w(\rvx{Si},\rvx{O}^4)$ is always one and can be omitted from
eq.~\ref{eqn:SiO2_ins_gen}.

The probability of generating a SiO$_2$ removal move is
\begin{eqnarray}
\label{eqn:SiO2_rem_gen}
\alpha[\mathrm{n}_{\ns+1,\no+2}\rightarrow\mathrm{o}_{\ns,\no}] 
&= &
	\frac{1}{n\left(\mathrm{Si}^{(2)}\right)^{\mathrm{(n)}}}
 \frac{1}{2}
\frac{k_1!}{{C_1}^{k_1}}\frac{\exp\left(-\beta{U_1}^{\mathrm{(o)}}\right)}{w_1^{\mathrm{(o)}}}\nonumber\\
&&\times\frac{(k_1-1)!}{{C_1}^{k_1-1}}\frac{(k_2-1)!}{(2\pi)^{k_2-1}}\,
\end{eqnarray}
where $w_1^{\mathrm{(o)}} = \sum_{k=0}^{k_1}\exp(-\beta
U_1^{(k)})$. Note that $w_1^{\mathrm{(n)}}$ and
$w_1^{\mathrm{(o)}}$ are asymmetric because in the SiO$_2$ removal move
$p(r_{\mathrm{OO}})/(4 \pi {r_{\mathrm{OO}}}^2)$
 is not used to bias the selection process.  The
first term at the right hand side accounts for the probability that we
choose the appropriate 2-connected Si
atom at random. We always choose the two O$^{(1)}$ atoms bonded to the chosen
Si$^{(2)}$ atom.  
The factor of $1/2$ accounts for the two possible choices for labeling
groups 1 and 2.
Substituting eqs.~\ref{eqn:p_states},
\ref{eqn:SiO2_ins_gen}, and \ref{eqn:SiO2_rem_gen} into
eq.~\ref{eqn:detailed_balance}, and using
${U_2}^{\mathrm{(n)}}+{U_1}^{\mathrm{(n)}}-{U_1}^{\mathrm{(o)}}=U^{\mathrm{(n)}}-U^{\mathrm{(o)}}$,
we obtain eq.~\ref{eqn:SiO2_acc} and thus complete the proof.

\section{The Reactive Moves}
\label{app:reactive}
In this section we describe the other Monte Carlo reactive moves.  A
SiO$_3$ insertion move proceeds as follows:
\begin{enumerate}
\item
Pick an O$^{(1)}$ atom from the cluster at random.
\item
Generate a monomeric SiO$_4$ configuration from the Boltzmann
distribution at the desired temperature. This is done by a short
hybrid Monte Carlo run.
\item
Take the SiO$_3$ configuration from the monomer by dropping the last O
atom, and graft this group onto the O$^{(1)}$ in step~1 in a way so that
the O atom left behind would fall on top of the the O$^{(1)}$. The
orientation of the fragment is chosen from a biased probability of the
cosine of the new Si-O-Si angle and the two associated torsional angles. 
Within the cluster and monomer partition functions, 
the two associated torsional angles and the orientation of the
monomer are already randomized.
We optionally choose to re-randomize these.
 The biased
probability is chosen to be $\exp(-\beta V_3)/C$, where $V_3$ is given
in eq.~\ref{eqn:v3} for the O atom and
\begin{eqnarray}
\label{eqn:angle_norm}
C &= &\int_{-1}^1 d(\cos\theta)\exp(-\beta V_3)
\end{eqnarray}
\end{enumerate}

A SiO$_3$ removal move picks a Si$^{(1)}$ atom at random and removes
the Si$^{(1)}$ atom and the three associated O$^{(1)}$ atoms from the
cluster.  The acceptance probability is given below:
\begin{eqnarray}
\label{eqn:SiO3_acc}
\frac{\mathrm{acc}[\mathrm{o}_{\ns,\no}\rightarrow\mathrm{n}_{\ns+1,\no+3}]}
     {\mathrm{acc}[\mathrm{n}_{\ns+1,\no+3}\rightarrow\mathrm{o}_{\ns,\no}]}
     & = & \zs\zo^3\frac{n\left(\mathrm{O}^{(1)}\right)^{\mathrm{(o)}}}
		{n\left(\mathrm{Si}^{(1)}\right)^{\mathrm{(n)}}}
               \frac{C^{\mathrm{(n)}}}{2}
		\frac{Z_{1,4}}{3! \exp(-\beta U_{\mathrm{m}})}\nonumber\\
&&		\times\exp[-\beta(\Delta U-{V_3}^{\mathrm{(n)}})],
\end{eqnarray}
where $\Delta U$ is the change of system energy, and $U_{\mathrm{m}}$
is the energy of the monomer.  The factor of $1/2$ accounts
for the fact that the monomer orientation is already random 
in the absence of a bias; equivalently, this factor accounts for the
values of $\cos \theta$ that contribute to the flux.
  The factor of 3! accounts for the
number of indistinguishable SiO$_3$ configurations that contribute 
to the flux.

The idea of the hydrolysis/condensation move is to couple an O atom
insertion/removal move to a local re-orientation of the nearest
atoms, so as to alleviate part of the energy barrier present along the reaction
path. We demonstrate in Fig.~\ref{fig:rxn21} how a typical
hydrolysis/condensation move alters the local configuration of a cluster.
The hydrolysis move proceeds as follows:
\begin{enumerate}
\item
\label{itm:pick_Si2}
Pick randomly a Si-O$^{(2)}$ bond that is to be broken later. If when
this bond is removed, the cluster breaks into two independent
clusters, reject this move. Otherwise, mark the two Si atoms to which this
O$^{(2)}$ is bonded. The ``first'' Si atom is the one in the chosen Si-O$^{(2)}$
bond. We name the O$^{(2)}$ atom in the Si-O$^{(2)}$ bond the pivot O
atom. Count the O$^{(2)}$ connectivities of the two Si atoms after
the Si-O bond has been broken.  Define the two connectivities to be $i_1$
and $i_2$, respectively. Denote as group 1 the first Si and the
$(4-i_1)$ O$^{(1)}$ atoms bonded to it. Denote as group 2 the second
Si and the $(4-i_2)$ O$^{(1)}$ atoms bonded to it.
\item
\label{itm:hd_g1}
We seek possible low energy positions for the O atom to be inserted by
defining an insertion volume around the first Si atom. The way this
volume is defined is quite arbitrary. This insertion bias affects the
efficiency of the moves but does not alter the equilibrium results. We
pick the insertion volume $V_{\mathrm{ins}}$ according to $i_1$:
\begin{itemize}
\item
If $i_1=3$, the insertion volume is defined as a sphere
centered at the pivot O atom with a radius $R_{\mathrm{c}}$. Therefore,
\begin{eqnarray}
	V_{\mathrm{ins}} &= & \frac{4}{3}\pi{R_{\mathrm{c}}}^3.
\end{eqnarray}
\item
If $i_1=2$, the insertion volume is a torus defined by revolving a
circle around an axis passing through the two O$^{(2)}$ atoms bonded
to the first Si atom. The circle resides on the plane defined by the two
O$^{(2)}$ atoms and the pivot O atom, is centered at the pivot O atom,
and has a radius $R_{\mathrm{c}}$. We find that for this choice
\begin{eqnarray}
V_{\mathrm{ins}} &= & 2\pi^2 r_{\mathrm{rev}}{R_{\mathrm{c}}}^2,
\end{eqnarray}
where $r_{\mathrm{rev}}$ is the distance between the pivot O atom and
the axis of revolution.
\item
If $i_1=1$, the volume is defined as a shell formed by two concentric
spheres centered at the only O$^{(2)}$ atom bonded to the first Si atom,
each having a radius of $r_{\mathrm{d}}+R_{\mathrm{c}}$ and
$r_{\mathrm{d}}-R_{\mathrm{c}}$, respectively. Here $r_{\mathrm{d}}$
is the distance between the O$^{(2)}$ atom and the pivot O atom. The
volume is
\begin{eqnarray}
	V_{\mathrm{ins}} &= & \frac{8}{3}\pi R_{\mathrm{c}}
			 (3{r_{\mathrm{d}}}^2+ {R_{\mathrm{c}}}^2).
\end{eqnarray}
\end{itemize}
Note that the radius $R_{\mathrm{c}}$ can be chosen independently in
different cases. We use $R_{\mathrm{c}}=0.3$~\AA\ in all cases. Now,
insert the new O atom randomly in the volume $V_{\mathrm{ins}}$, and move
group 1 as a rigid body, so that the Si atom in group 1 moves away from
the pivot O atom and bonds to the inserted O atom.  This move is
done according to the position of the insertion O atom:
\begin{itemize}
\item
If $i_1=3$, do not move group 1.
\item
If $i_1=2$, rotate group 1 around an axis passing through the two
O$^{(2)}$ atoms bonded to the Si atom, so that the plane
defined by the two O$^{(2)}$ atoms and the pivot O atom, when rotated,
will coincide with the plane defined by the two O$^{(2)}$ atoms and
the inserted O atom.
\item
If $i_1=1$, rotate group 1 so that the axis passing through the
O$^{(2)}$ atom and the pivot O atom, when rotated, will coincide with
the axis passing through the O$^{(2)}$ atom and the inserted O atom.
This is followed by a random rotation of group 1 around the
axis.
\end{itemize}
\item
\label{itm:hd_g2}
Next, move group 2 according to the connectivity $i_2$:
\begin{itemize}
\item
If $i_2=3$, do not move the Si atom. Choose randomly
a new position for the pivot O atom from a sphere centered at the
pivot O atom and having a radius $R_{\mathrm{c}}$.
\item 
If $i_2=2$, rotate group 2 by a random angle around the axis
passing through the two O$^{(2)}$ atoms bonded to the second Si atom.  
\item 
If $i_2=1$, choose a new orientation for group 2 from a uniform
distribution of the cosine of the Si-O$^{(2)}$-Si bond angle 
and the two neighboring torsional angles.
\end{itemize}
\item
Configurational bias Monte Carlo is used in step~\ref{itm:hd_g1} and
step~\ref{itm:hd_g2} to enhance the acceptance probability.  The trial
positions for the inserted O and trial moves for group 2 are generated
$k$ times, where $k$ is an positive integer. We use $k=100$. Each set of positions is
picked with a probability proportional to the $\exp(-\beta U)$, $U$
being the energy of the configuration. Calculate the Rosenbluth factor
for these trial moves.\cite{Frenkel_96}
\end{enumerate}

A condensation move attempts to remove an O$^{(1)}$ atom
and alter the cluster configuration so that the Si atom that 
lost an O atom becomes bonded to another O atom. The condensation move
proceeds as follows:
\begin{enumerate}
\item
Pick randomly a pair of O$^{(1)}$ atoms from a condensation list that
is constantly updated during the simulation.  The two O$^{(1)}$ atoms
must not be bonded to the same Si atom, and the two Si atoms to which
the two O$^{(1)}$ atoms are bonded must not be bonded to the same
O$^{(2)}$. The latter restriction avoids cyclic 2-Si rings.  The
two O atoms must also satisfy certain geometrical criteria that are
used to eliminate geometrically impossible pairs so as to improve
efficiency of the move. However, the list must include all pairs that
can be reached by the reverse move, i.e., the hydrolysis move. In the
condensation list the O atom pairs $(i,j)$ and $(j,i)$ are counted as
two different pairs, while in the SiO$_2$ insertion list the O atoms
are counted as one pair. Denote the Si atom bonded to the first O atom
and the associated O$^{(1)}$ atoms as group 1, and denote the Si atom bonded
to the second O atom and the associated O$^{(1)}$ atoms as group
2. Name the second O$^{(1)}$ atom from the list the pivot atom.
\item 
Move group 2 atoms in the same way as described in step~\ref{itm:hd_g2}
of the hydrolysis move.
\item
Remove the first O$^{(1)}$ atom from group 1. 
Reorient group 1 so as to form a new bond between the Si atom of group 1 and the pivot
atom. This reorientation depends on the connectivity and is the same
as step~\ref{itm:hd_g1} of the hydrolysis
move, except that the roles of the pivot atom and the inserted atom are
reversed. Multiple trial moves are generated, and the
Rosenbluth factor for these trial moves are calculated.
\end{enumerate}
The acceptance probability is
\begin{eqnarray}
\label{eqn:hd/cn_acc}
\frac{\mathrm{acc}[\mathrm{o}_{\ns,\no}\rightarrow\mathrm{n}_{\ns,\no+1}]}
     {\mathrm{acc}[\mathrm{n}_{\ns,\no+1}\rightarrow\mathrm{o}_{\ns,\no}]}
     & = & \zo\frac{n(\mathrm{SiO})^{\mathrm{(o)}}}
		{n(\mathrm{OO})^{\mathrm{(n)}}}
	   \frac{w^{\mathrm{(n)}}}{w^{\mathrm{(o)}}}V_{\mathrm{ins}},
\end{eqnarray}
where $n(\mathrm{SiO})$ and $n(\mathrm{OO})$ are the number of
qualified Si-O and O-O pairs, respectively, and $w^{\mathrm{(n)}}$
and $w^{\mathrm{(o)}}$ represent the Rosenbluth factors for the forward and
reverse moves, respectively.

A prune-and-graft move attempts to ``prune'' a fragment off the
cluster and ``graft'' it onto a different site of the cluster.  It
proceeds as follows:
\begin{enumerate}
\item
Pick a Si-O$^{(2)}$ bond at random. If the cluster stays a cluster when
this bond is removed, then we reject this move because the branch so generated
is not free. Otherwise, mark the entire fragment that is
still connected to this O$^{(2)}$ after cleavage of the Si-O$^{(2)}$
bond.
\item
Pick randomly an O$^{(1)}$ that is not part of the marked fragment.
\item
Move the marked fragment in a way for the Si atom liberated in step~1 to bond to
the O$^{(1)}$ chosen in step~2. During the move, the fragment is treated as a
rigid body. Require
that the bond has the same length as the bond broken in step~1.
The orientation of the fragment is chosen from a biased distribution
of the new Si-O-Si angle and from a uniform distribution of the two
neighboring torsional angles. The biasing probability is the same as
that used in SiO$_3$ insertion move, with the normalization constant
$C$ given by eq.~\ref{eqn:angle_norm}.
\end{enumerate}

The acceptance probability is
\begin{eqnarray}
\frac{\mathrm{acc}(\mathrm{o}_{\ns,\no}\rightarrow\mathrm{n}_{\ns,\no})}
	{\mathrm{acc}(\mathrm{n}_{\ns,\no}\rightarrow\mathrm{o}_{\ns,\no})}
 &= & \frac{C^{\mathrm{(n)}}\exp(-\beta U^{\mathrm{(n)}})}
	{C^{\mathrm{(o)}}\exp(-\beta U^{\mathrm{(o)}})},
\end{eqnarray}

\section{Computing the Configurational Integral
	 of a Monomer}
\label{app:monomer_p}
The configurational integral for a monomer, $Z_{1,4}$, is calculated
by thermodynamic integration from infinite temperature to the
synthesis temperature:
\begin{eqnarray}
\label{eqn:mono_int}
\ln Z_{1,4} &= & \ln Z_{1,4}(\beta=0) - \int_0^\beta \langle
	 		U_{\mathrm{m}} \rangle_{\beta'}\, d\beta'.
\end{eqnarray}
Here $\langle U_{\mathrm{m}} \rangle_{\beta'}$ is the canonical ensemble average
of the monomer energy at $\beta'$, and
\begin{eqnarray}
Z_{1,4}(\beta=0) &= & \int
d\rvx{Si}d\rvx{O}\delta(\mathbf{R}_{\mathrm{cm}})w(\rvx{Si},
\rvx{O}^4)\nonumber \\
& = &\int
d\xv{1}d\xv{2}d\xv{3}d\xv{4}d\mathbf{R}_{\mathrm{cm}} w(\mathbf{0},
\xv{}^4)\delta(\mathbf{R}_{\mathrm{cm}})
\nonumber\\
& =&\left(\frac{4}{3}\pi {R_{\mathrm{max}}}^3\right)^4,
\end{eqnarray}
where $\xv{i}=\rv{\mathrm{O}i}-\rvx{Si}$, and $R_{\mathrm{max}}=2.6$
\AA\ is the maximum Si-O bond length.  We compute the value of this
integral by splitting the range of integration into three regions, and
we use 16-point Gauss-Legendre integration in each region.  Performing
the integration of eq.~\ref{eqn:mono_int}, we find $Z_{1,4} =
\exp(311.31)$ \AA$^{12}$ at $T=430$~K. An alternative way of computing
the free energy is to couple the cluster to a set of non-interacting
harmonic oscillators whose free energy is known, using a parameter
$\lambda$ to adjust the relative contributions of the two
systems. Integration from $\lambda=0$ to $\lambda=1$ gives the free
energy difference between the monomer and the harmonic oscillators
and, thus, the absolute free energy of the cluster.  To compute the
integral, we again split the range of integration into three regions
and use 16-point Gauss-Legendre integration.  The value obtained in
this way agrees with the above result within statistical error, $\ln
Z_{1,4}=311.12$.  Note that in the harmonic oscillator reference
system, the atoms are distinguishable, and so an ``identity exchange''
move must be added to the simulation so that the free energy
difference between the indistinguishable cluster and the
distinguishable reference system is properly calculated.  This
distinguishability factor, as well as the rotational entropy, shows up
as a large contribution at small $\lambda$ in the integral for the
free energy difference.  We adopt the Einstein crystal result, $\ln
Z_{1,4}=311.12$, as it is likely more accurate because the thermodynamic
integration that gives this result avoids the singularity that exists
at $\beta=0$ in eq.\ \ref{eqn:mono_int}.

To test detailed balance for our SiO$_3$ insertion and removal moves,
we compared the values of $\ln (Z_{2,7}/Z_{1,4})=311.88$ 
obtained from our simulation using only the SiO$_3$ insertion and
removal moves to change the number of atoms with that from the
integration via harmonic oscillators, $\ln (Z_{2,7}/Z_{1,4})=311.77$.
The results agree within statistical errors.  To test detailed balance
for the hydrolosis and condensation moves, we compared the free energy
differences between a liner and cyclic tri-silicate species.  We find
the result $\ln (Z_{3,10}/Z_{3,9}) = 6.32$
 from a simulation employing only the hydrolosis and condensation
moves to change the number of atoms, and we find $\ln
(Z_{3,10}/Z_{3,9}) = 6.38$
 from the integration via harmonic oscillators.  The level of
agreement is, again, within statistical errors.  To check the SiO$_2$
insertion and removal move, we consider the equilibration between a
dimer and a cyclic trimer.  From the simulation, we find $\ln
(Z_{3,9}/Z_{2,7}) = 307.84$.
From the integration via harmonic oscillators we find $\ln
(Z_{3,9}/Z_{2,7}) = 307.65$.
Again, the results agree within statistical errors.  Finally, we check
a combination of the hydrolosis and condensation moves and the SiO$_2$
insertion and removal moves.  From the above thermodynamic integration
results, we find $\ln (Z_{3,10}/Z_{2,7}) = 314.03$.  From simulation,
we find $\ln (Z_{3,10}/Z_{2,7}) = 313.96$, which is again within
statistical errors.

\bibliography{nucleation}

\clearpage

\begin{table}[tbp]
\caption{Constants in the interaction potential for silicate ions. The
units of length and energy are \AA\ and $e^2$/\AA$=14.39$ eV,
respectively.}
\label{tab:potential}
\begin{center}
\begin{tabular}{ccccc}
	&	q (e)	&	$\alpha$ (\AA$^3$)\\ 
\hline
Si	&	1.60	&	0.00    \\
O       &	-0.80 (for O$^{(2)}$)	&	2.40	\\
        & 	-0.90 (for O$^{(1)}$)   &		\\[.1in]
	&	$\eta$	&	$H$	\\
\hline
Si-Si	&	11	&	$\ $0.057	\\
Si-O	&	9	&	11.387	\\
O-O	&	7	&	51.692	\\[.1in]
	&	$B$	&	$l$	&$\overline{\theta}$ (deg)&	$r_0$	\\
\hline
Si-O-Si &	1.40	&	1.0	&	141.00	&	2.60	\\
O-Si-O	&	0.35	&	1.0	&	109.47	&	2.60	\\
\end{tabular}
\end{center}
\end{table}
\begin{table}[tbp]
\caption{The nucleation barrier and critical cluster size at different
Si concentration and pH = 12.
 At 0.017~M, the free energy of cluster does not pass a
maximum up to $\ns=200$; thus the nucleation barrier and critical
cluster are undefined.}
\label{tab:barrier}
\begin{center}
\begin{tabular}{ccc}
Si concentration (M) & $\beta\Delta\Omega$ (dimensionless) & Critical Cluster ($\ns$)	\\ 
\hline
0.75$\ $  	& $\ $ 66.2 & 27 \\
0.33$\ $	& $\ $ 91.6 & 32 \\
0.08$\ $	& 126.8  & 45   \\
0.017	& -	& -
\end{tabular}
\end{center}
\end{table}

\begin{flushleft}

\begin{figure}[p]
\caption{The reactive moves change the number of atoms, as shown on
the $\ns$, $\no$ lattice. The two dashed lines, $\no=3\ns+1$ and $\no=2\ns$,
represent the upper and lower bounds for the possible lattice points.}
\label{fig:lattice}
\end{figure}

\begin{figure}[h]
\caption{A SiO$_2$ insertion/removal move. The upward
arrow represents an insertion move, while the downward arrow represents
a removal move.}
\label{fig:rxnSiO2}
\end{figure}

\begin{figure}[h]
\caption{The connectivities of Si atoms for different values of
$\ns$. The statistical errors are smaller than the size of the
points.\label{fig:connect}}
\end{figure}

\begin{figure}[h]
\caption{A snapshot of a cluster with
$\ns=77$.\hfill\label{fig:ns_77}}
\end{figure}

\begin{figure}[h]
\caption{The probability distributions for the adjacent T-T atom
distances, T being Si or Al. Note that the cluster is simulated at
$T=430$~K, while the other two distributions are computed from
crystal structures.\label{fig:T-T}}
\end{figure}

\begin{figure}[h]
\caption{The probability distributions for the T-T-T angles, T being
Si or Al.\hfill\label{fig:T-T-T}}
\end{figure}

\begin{figure}[h]
\caption{a) Average ring size distribution at $\ns=77$ and
$\ns=200$. b) Ring size distribution of known zeolite structures.
\label{fig:ring_size}}
\end{figure}

\begin{figure}[h]
\caption{Probability of observing the centroid of a
3-ring at a given radius from the center of the
cluster for a 197 Si-atom cluster.  The distance is
scaled by the radius of gyration of the cluster, $R_{\rm g}$.
\label{fig:3ring}}
\end{figure}

\begin{figure}[h]
\caption{The free energy of cluster formation along the nucleation
coordinate at pH = 12. The unit of the Si monomer concentration, $\rhos$, is M
(mol/$l$). The dashed line represents the free energy of cluster
formation at
$\rhos=0.33$ M, beyond the critical cluster size. The free energies of
quartz cluster formation are calculated using values of $\mus$ and $\muo$
corresponding to $\rhos=0.33$ M and pH = 12. The statistical errors are roughly
the size of the points.}
\label{fig:cl_g}
\end{figure}

\begin{figure}[h]
\caption{The free energy of cluster formation along the nucleation
coordinate at different values of the pH and a fixed silicon monomer
concentration of $\rhos = 0.33$ M. }
\label{fig:cl_g_pH}
\end{figure}

\begin{figure}[h]
\caption{A typical hydrolysis/condensation reactive move. The downward
arrow represents a hydrolysis move, while the upward arrow represents
a condensation move.}
\label{fig:rxn21}
\end{figure}

\clearpage
\newpage
\begin{center}

\includegraphics[width=4in]{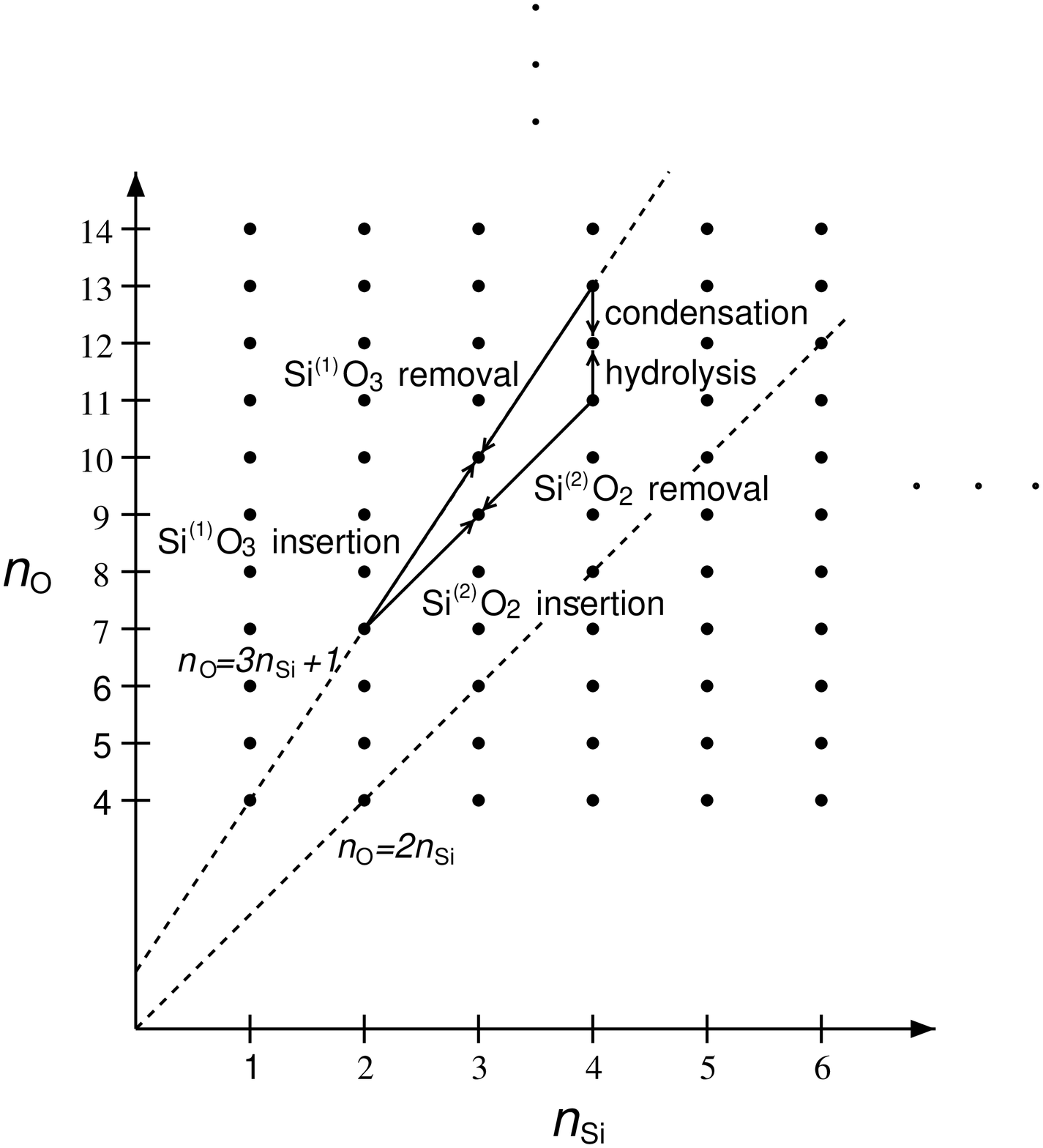}

\vspace{1.5in}
Figure \ref{fig:lattice}
 Wu and Deem, ``Monte Carlo Study\ldots.''

\newpage
\includegraphics[height=4in]{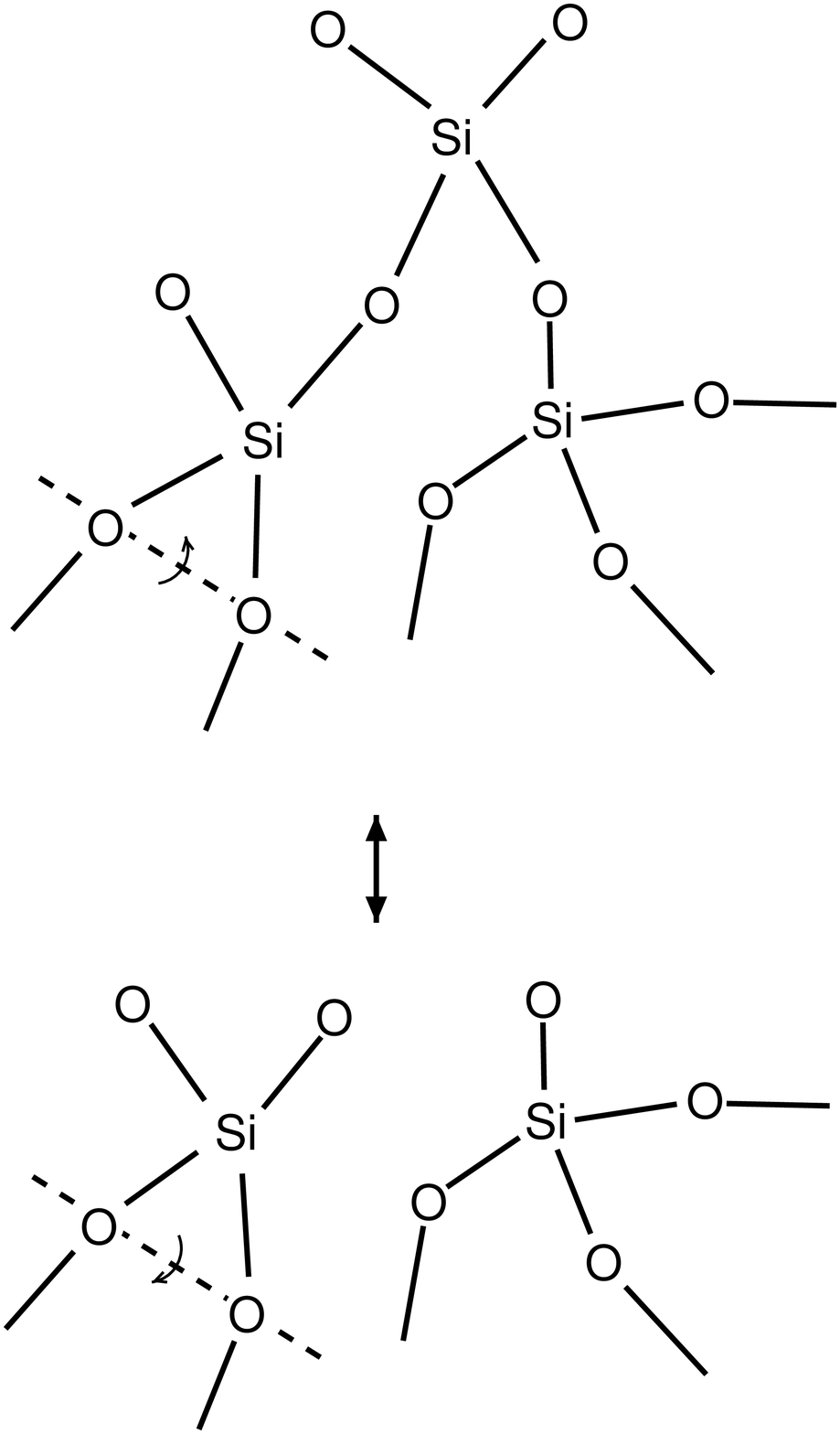}

\vspace{1.5in}
Figure \ref{fig:rxnSiO2}
 Wu and Deem, ``Monte Carlo Study\ldots.''

\newpage
\includegraphics[width=4in]{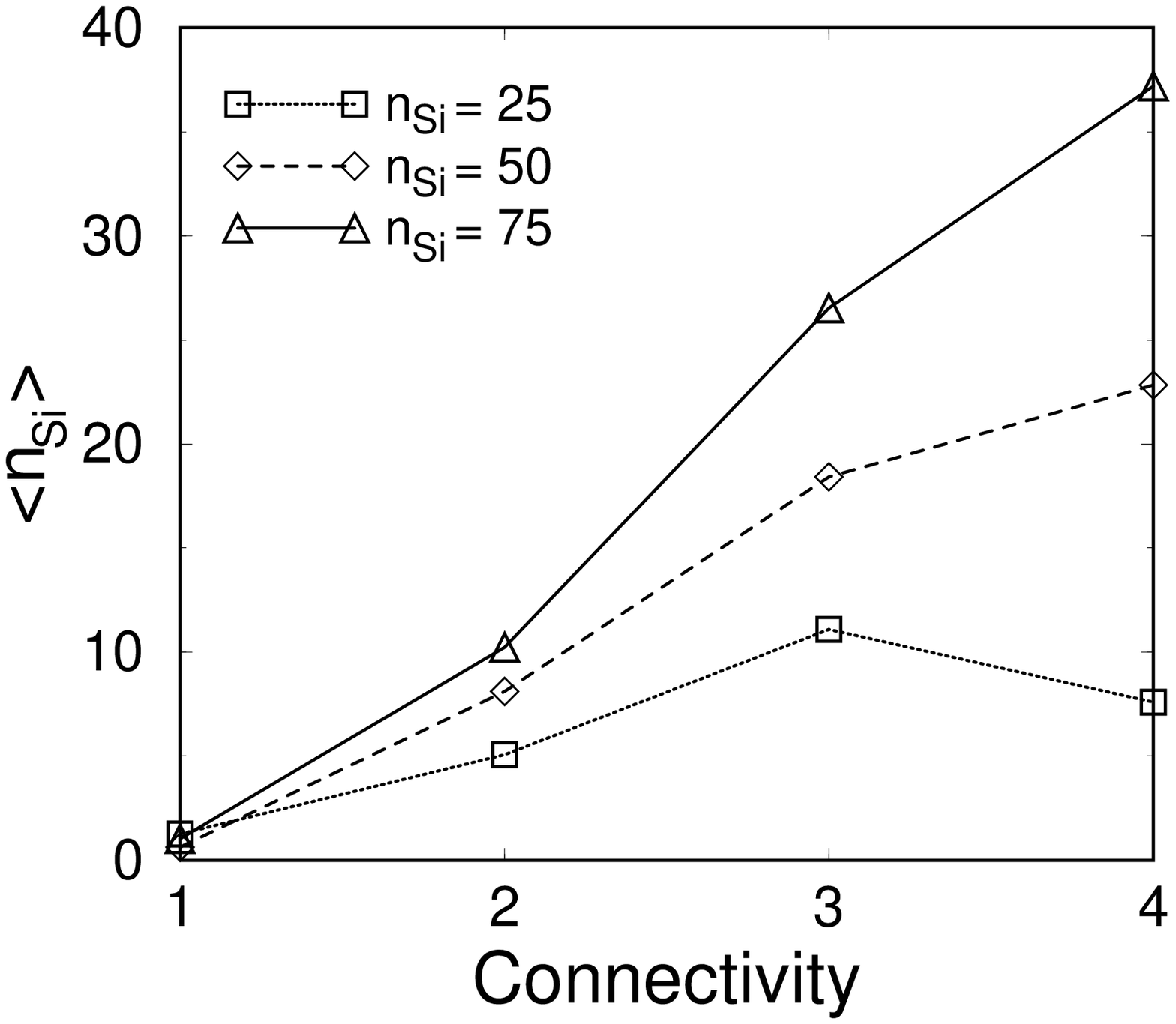}

\vspace{1.5in}
Figure \ref{fig:connect}
 Wu and Deem, ``Monte Carlo Study\ldots.''

\newpage
\includegraphics[width=4in]{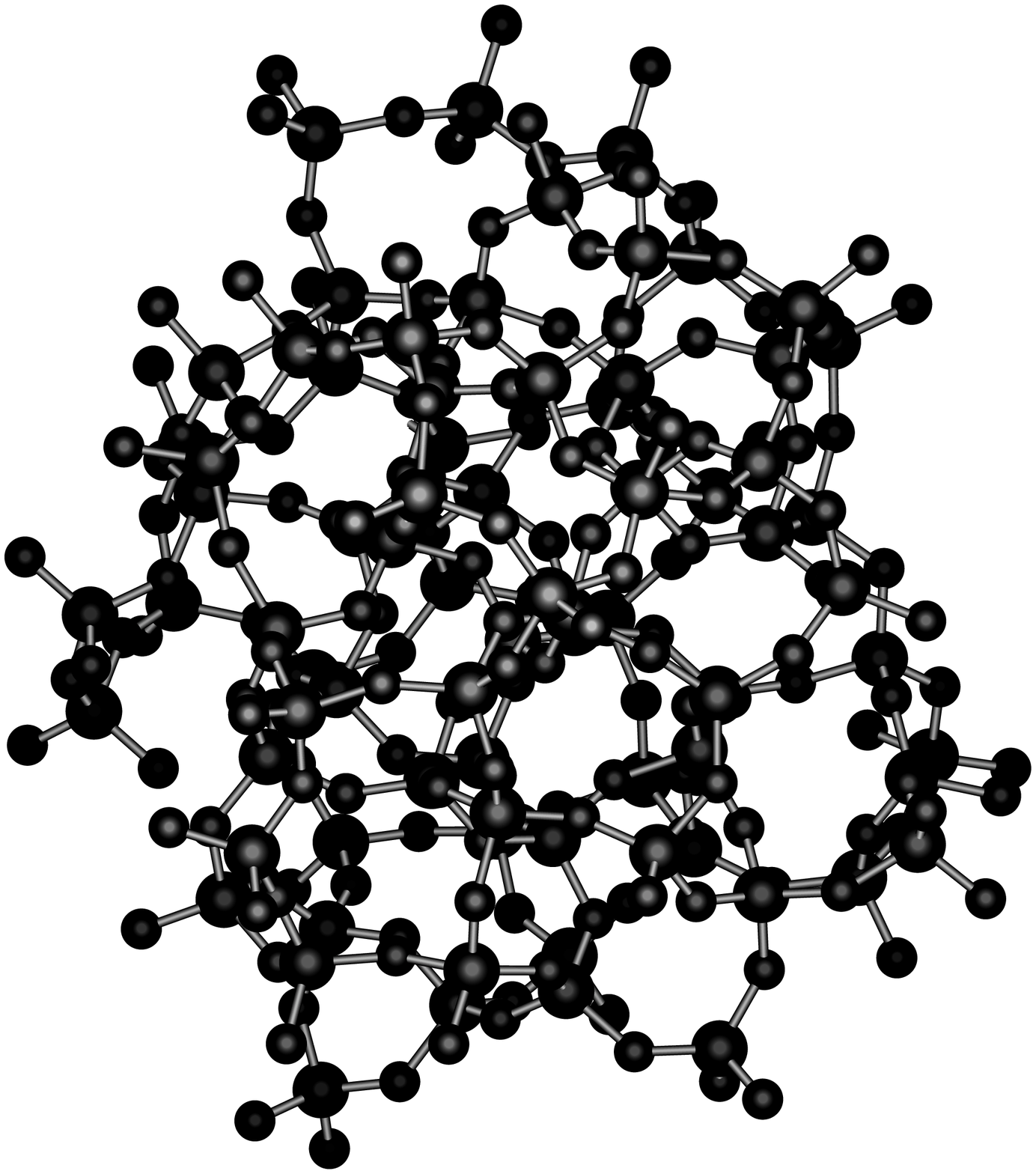}

\vspace{1.5in}
Figure \ref{fig:ns_77}
 Wu and Deem, ``Monte Carlo Study\ldots.''

\newpage
\includegraphics[width=4in]{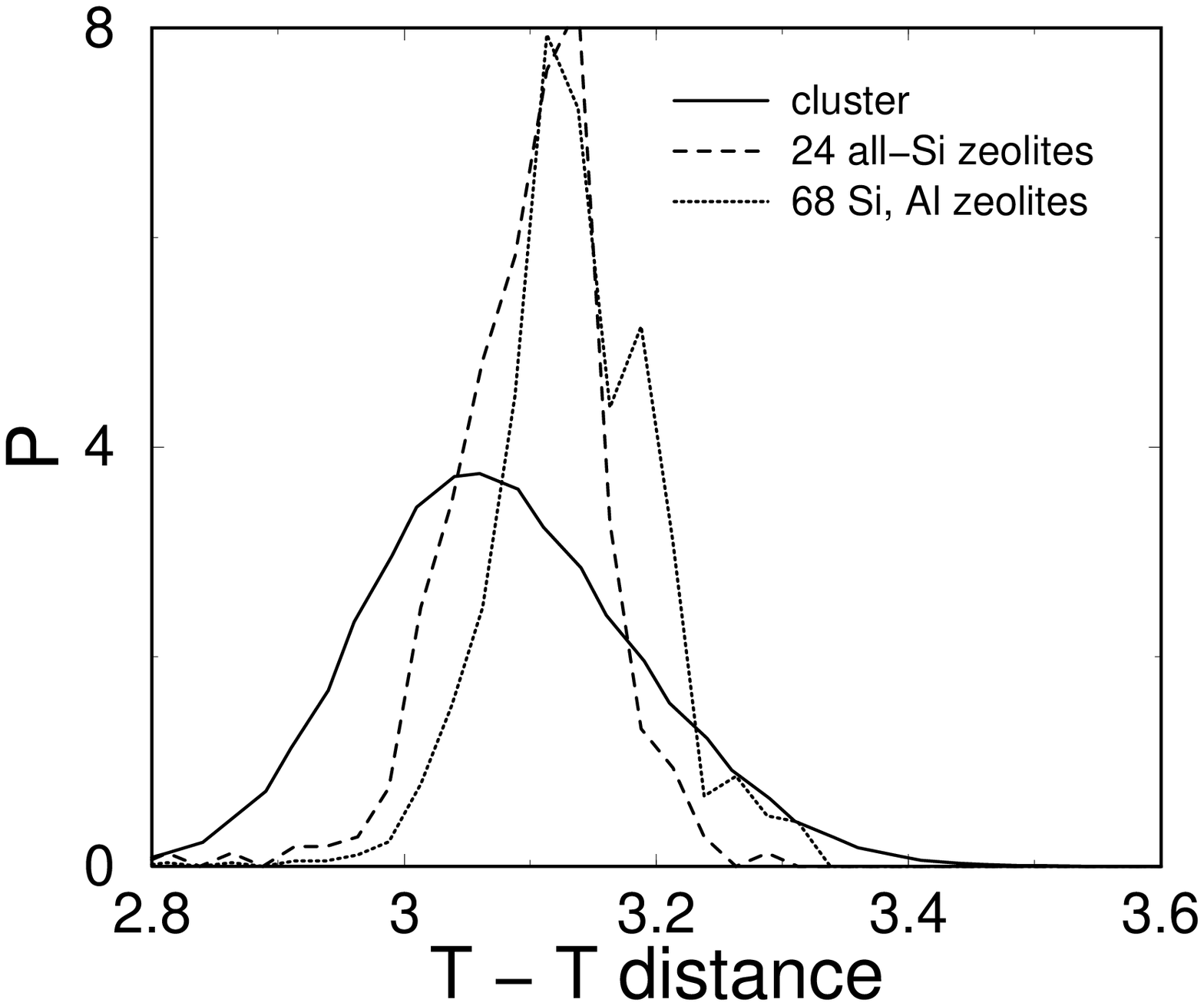}

\vspace{1.5in}
Figure \ref{fig:T-T}
 Wu and Deem, ``Monte Carlo Study\ldots.''

\newpage
\includegraphics[width=4in]{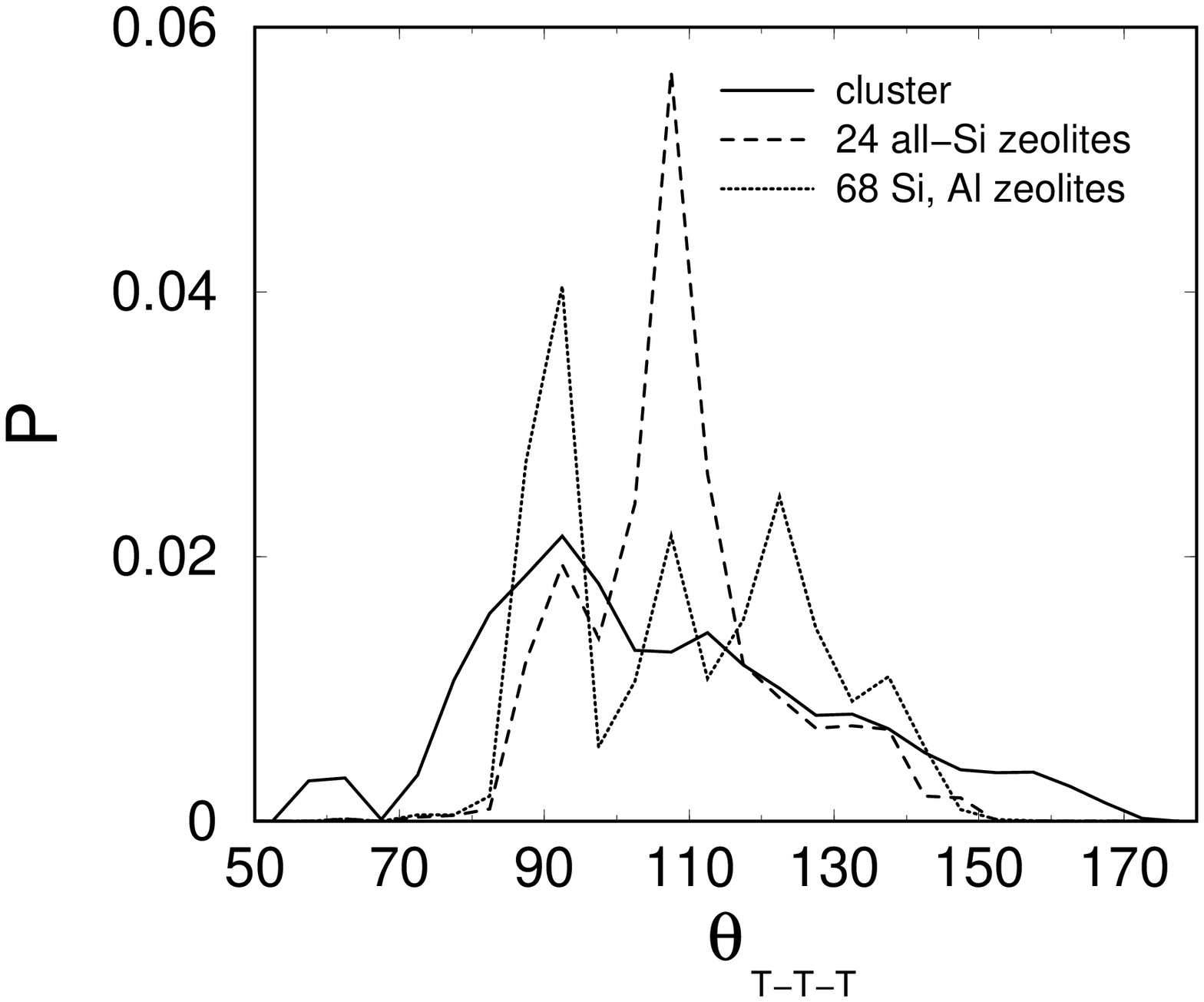}

\vspace{1.5in}
Figure \ref{fig:T-T-T}
 Wu and Deem, ``Monte Carlo Study\ldots.''

\newpage
\includegraphics[width=3in]{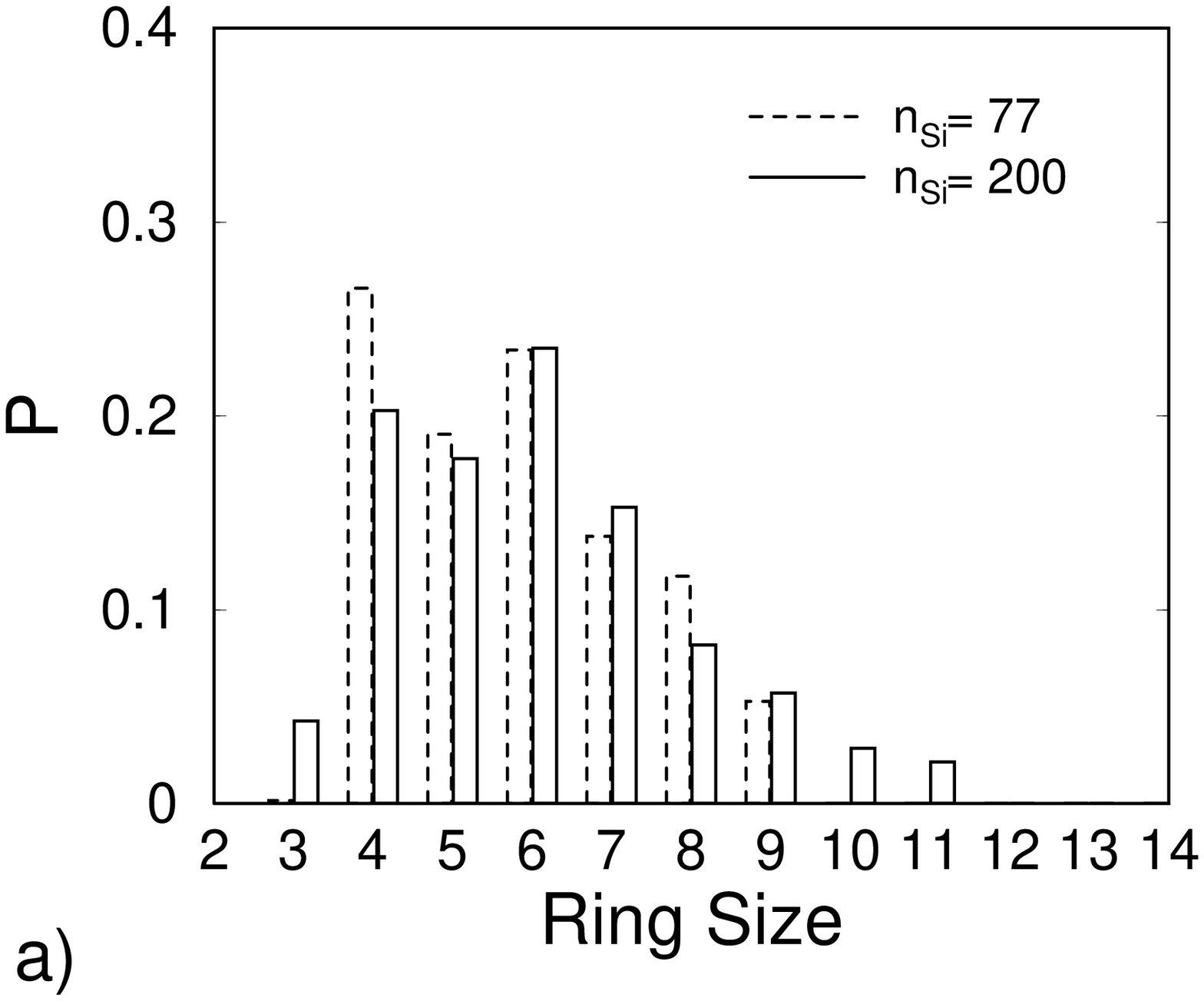}
\hfill
\includegraphics[width=3in]{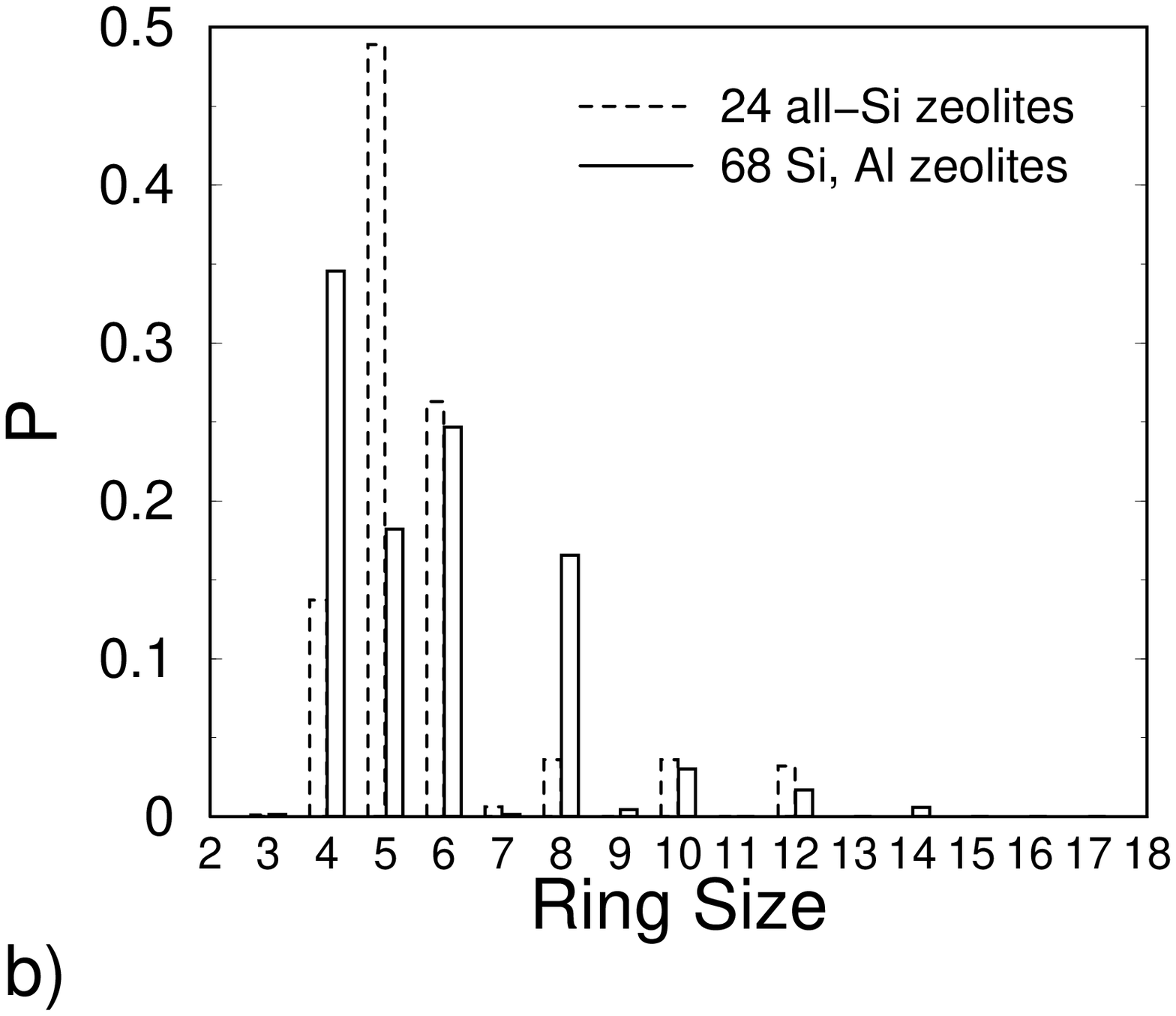}

\vspace{1.5in}
Figure \ref{fig:ring_size},
 Wu and Deem, ``Monte Carlo Study\ldots.''

\newpage
\includegraphics[width=4in,clip=]{fig8.eps}

\vspace{1.5in}
Figure \ref{fig:3ring}
 Wu and Deem, ``Monte Carlo Study\ldots.''

\newpage
\includegraphics[width=4in]{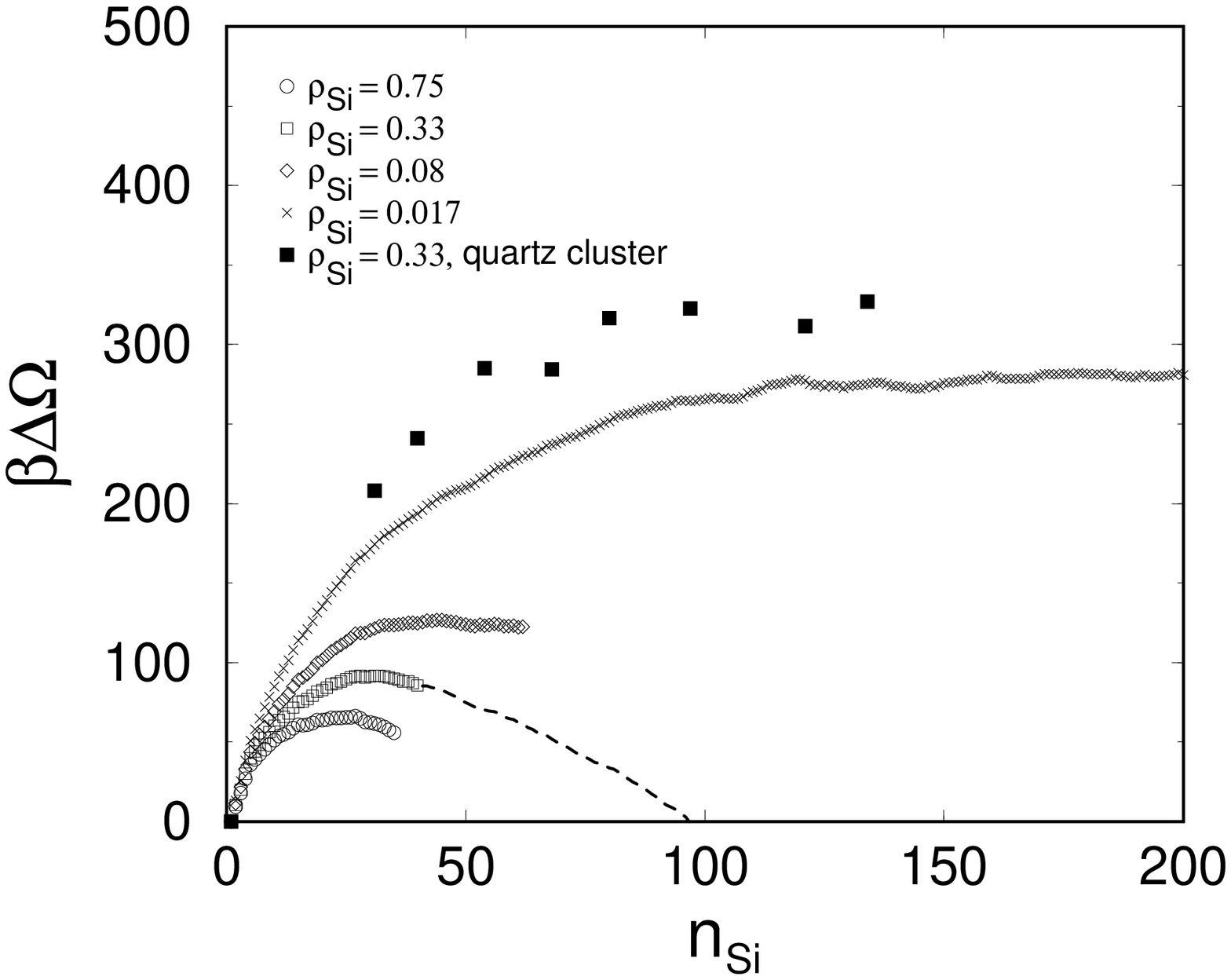}

\vspace{1.5in}
Figure \ref{fig:cl_g}
 Wu and Deem, ``Monte Carlo Study\ldots.''

\newpage
\includegraphics[width=4in]{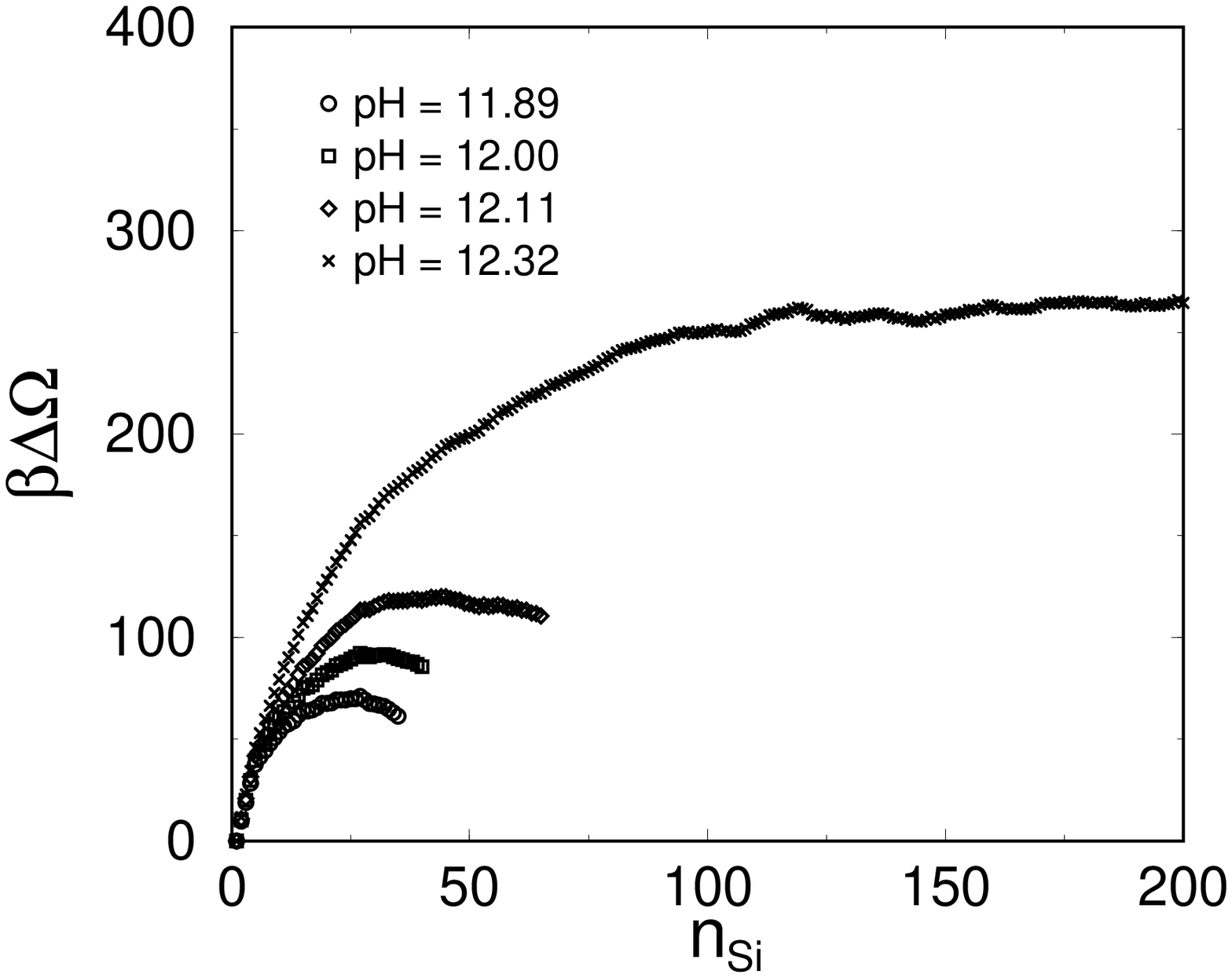}

\vspace{1.5in}
Figure \ref{fig:cl_g_pH}
 Wu and Deem, ``Monte Carlo Study\ldots.''

\newpage
\includegraphics[width=4in]{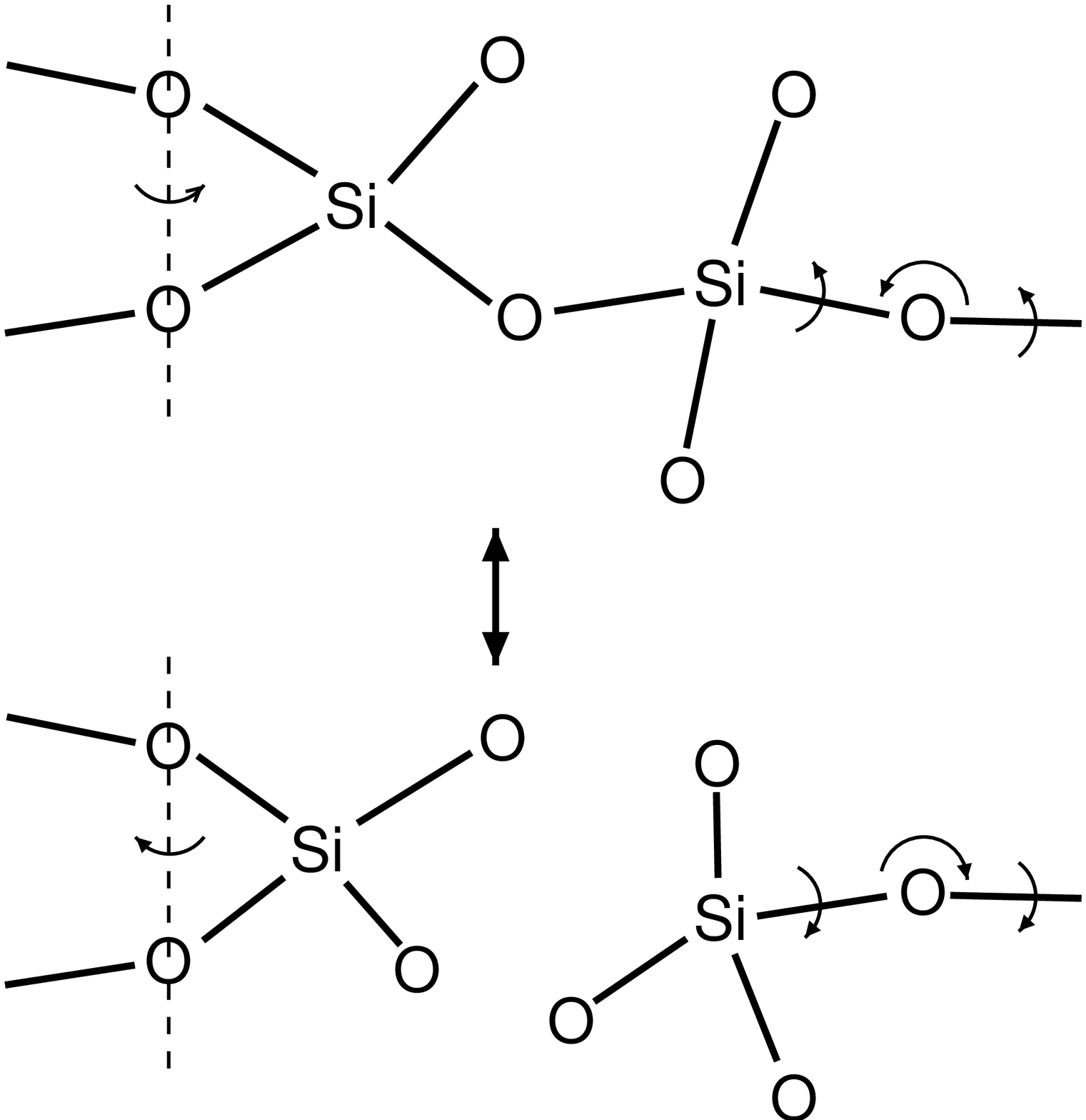}

\vspace{1.5in}
Figure \ref{fig:rxn21}
 Wu and Deem, ``Monte Carlo Study\ldots.''

\end{center}
\end{flushleft}

\end{document}